\documentclass[twocolumn,trackchanges]{aastex63}

\usepackage{amsmath}
\usepackage{graphicx}
\usepackage{soul}

\newcommand{\chianti}{\textsc{chianti}}
\newcommand{\ramses}{\textsc{ramses}}
\newcommand{\bpass}{\textsc{bpass}}
\newcommand{\ramsesrt}{\textsc{ramses-rt}}

\newcommand{\yt}{\textsc{yt}}

\definecolor{red}{RGB}{250,0,0}
\definecolor{magenta}{RGB}{167,0,135}

\definecolor{pink}{RGB}{255,51,153}

\definecolor{green}{RGB}{0,153,0}
\definecolor{blue}{RGB}{0,0,250}

\definecolor{brown}{RGB}{150,74,0}

\setstcolor{red}

\newcommand{\nowind}{\texttt{NoWind\_rich}}
\newcommand{\rich}{\texttt{FaceWind10\_rich}}
\newcommand{\normal}{\texttt{FaceWind10}}
\newcommand{\myr}{\,{\rm Myr}}

\newcommand{\msun}{\mbox{$\rm M_{\odot}$}}
\newcommand{\msunyr}{\mbox{$\rm M_{\odot}\,{\rm yr^{-1}}$}}
\newcommand{\kms}{\mbox{${\rm km\,s^{-1}}$}}

\newcommand{\cmq}{\mbox{$\,{\rm cm^{-3}}$}}
\newcommand{\nH}{\mbox{$n_{\rm H}$}}
\newcommand{\SBR}{\mbox{$F_{\rm X}/F_{\rm H\alpha}$}}

\defcitealias{lee20}{L20}

\received{\today}
\revised{\today}

\shorttitle{Simulating Jellyfish Galaxies}
\shortauthors{Jaehyun Lee et al.}

\begin{document}
\title{Simulating Jellyfish Galaxies: A Case Study for a Gas-Rich Dwarf Galaxy}

\author[0000-0002-6810-1778]{Jaehyun Lee}
\affiliation{Korea Institute for Advanced Study, 85, Hoegi-ro, Dongdaemun-gu, Seoul 02455, Republic of Korea}

\author[0000-0002-3950-3997]{Taysun Kimm}
\affiliation{Department of Astronomy, Yonsei University, 50 Yonsei-ro, Seodaemun-gu, Seoul 03722, Republic of Korea}

\author[0000-0002-7534-8314]{J\'er\'emy Blaizot}
\affiliation{Univ Lyon, Univ Lyon1, Ens de Lyon, CNRS, Centre de Recherche Astrophysique de Lyon UMR5574, F-69230 Saint-Genis-Laval, France}

\author{Harley Katz}
\affiliation{Astrophysics, University of Oxford, Denys Wilkinson Building, Keble Road, Oxford OX1 3RH, UK}

\author[0000-0002-1566-5094]{Wonki Lee}
\affiliation{Department of Astronomy, Yonsei University, 50 Yonsei-ro, Seodaemun-gu, Seoul 03722, Republic of Korea}

\author{Yun-Kyeong Sheen}
\affiliation{Korea Astronomy and Space Science Institute, 776 Daedeokdae-ro, Yuseong-gu, Daejeon 34055, Republic of Korea}

\author[0000-0002-8140-0422]{Julien Devriendt}
\affiliation{Astrophysics, University of Oxford, Denys Wilkinson Building, Keble Road, Oxford OX1 3RH, UK}

\author{Adrianne Slyz}
\affiliation{Astrophysics, University of Oxford, Denys Wilkinson Building, Keble Road, Oxford OX1 3RH, UK}

\email{syncphy@gmail.com,tkimm@yonsei.ac.kr}

\begin{abstract}
 We investigate the formation of jellyfish galaxies using radiation-hydrodynamic simulations of gas-rich dwarf galaxies with a multi-phase interstellar medium (ISM).  We find that the ram-pressure-stripped (RPS) ISM is the dominant source of molecular clumps in the near wake within 10\,kpc from the galactic plane, while in-situ formation is the major channel for dense gas in the distant tail of the gas-rich galaxy. Only 20\% of the molecular clumps in the near wake originate from the intracluster medium (ICM); however, the fraction reaches 50\% in the clumps located at $80\,{\rm kpc}$ from the galactic center since the cooling time of the RPS gas tends to be short due to the ISM--ICM mixing ($\lesssim$ 10 Myr).  The tail region exhibits a star formation rate of $0.001-0.01\,\msunyr$, and most of the tail stars are born in the stripped wake within 10\,kpc from the galactic plane. These stars induce bright H$\alpha$ blobs in the tail, while H$\alpha$ tails fainter than $6\times10^{38}\,{\rm erg\,s^{-1}\,kpc^{-2}}$ are mostly formed via  collisional radiation and heating due to mixing. We also find that the stripped tails have intermediate X-ray to H$\alpha$ surface brightness ratios (1.5$\la \SBR \la$20), compared to the ISM ($\la$1.5) or pure ICM ($\gg$20). Our results suggest that jellyfish features emerge when the ISM from gas-rich galaxies is stripped by strong ram pressure, mixes with the ICM, and enhances the cooling in the tail.
\end{abstract}


\section{Introduction}

Ram pressure stripping, characterized by tails, is a key mechanism that accelerates galaxy evolution in cluster environments~\citep{gunn72,davies73,boselli06}. Observations have revealed multi-phase tails in HI~\citep{kenney04,oosterloo05,chung07,chung09,scott10,scott12,scott18}, H$\alpha$~\citep{gavazzi01,cortese06,cortese07,sun07,yagi07,yagi10,fumagalli14,Boselli16,poggianti17, sheen17}, and even in X-ray bands~\citep{finoguenov04,wang04,machacek05,sun05,sun06,sun10}, demonstrating that the interstellar medium (ISM) is efficiently removed from a galaxy by ram pressure and eventually dispersed by interactions with the intracluster medium (ICM). Although ram pressure stripping is believed to quench star formation on long time scales \citep{koopmann04a,koopmann04b}, various empirical and numerical studies have revealed other complicated effects of ram pressure stripping on galaxies \citep[e.g.,][]{grishin21,mun21}. For example, ram pressure can enhance star formation in satellite galaxies by compressing the ISM in the early stages of the infall \citep[e.g][]{steinhauser12,vulcani18}. Features associated with young stars are observed in some ram-pressure-stripped (RPS) wakes~\citep[e.g.][]{owers12,ebeling14,fumagalli14,rawle14,poggianti16,sheen17,jachym17,jachym19}, denoting the presence of dense molecular clouds surrounded by hot gas with temperatures higher than several million Kelvin. 

Indeed, extra-planar and tail CO emission is observed from RPS galaxies~\citep{jachym14,verdugo15,jachym17,lee17,lee18,moretti18,jachym19}. Some even have massive molecular clouds of $M_{\rm H_2}\sim10^9\,M_{\odot}$ in their tails~\citep{jachym17,moretti18,jachym19}, with a hint of star formation~\citep[see][for further details]{jachym17,jachym19}. Given that the lifetime of dense molecular clouds is typically less than $10\myr$ in an idealized environment without ICM winds \citep[e.g.][]{blitz80,vazquez-semadeni05} and the orbital velocity of a cluster satellite galaxy (i.e. wind velocity in wind tunnel experiments) is typically less than a few thousand $\rm km\,s^{-1}$, the molecular clouds observed tens of kpc away from the mid-plane of a galactic disk may have been formed {\it in-situ}, rather than directly stripped from the main body of a galaxy~\citep{jachym17}.

Many attempts have been made to understand the impact of ram pressure stripping on galaxies using numerical approaches. Disk stripping processes have been intensively examined for ICM winds with various properties~\citep{schulz01,vollmer01,roediger06a,vollmer06,roediger07,roediger08,jachym09}.  Furthermore, studies have explored the role of magnetic fields in ram pressure stripping~\citep{ruszkowski14,shin14,tonnesen14,ramos-martinez18}. Star formation in RPS disks and tails is another topic that has been investigated using numerical simulations \citep{schulz01,vollmer01,bekki03,kronberger08,kapferer08,kapferer09,steinhauser12,tonnesen12}. However, studies examining the complicated interplay between ram pressure and the multi-phase ISM driven by star formation and stellar feedback are lacking. \citet[][L20 hereafter]{lee20} investigated the impact of varying ICM winds on the multi-phase disk of a dwarf-sized galaxy using a suite of radiation-hydrodynamic (RHD) simulations. They showed that mild ICM winds with ram pressure $P_{\rm ram}/k_{\rm B}=5\times10^{4}\,{\rm K\,cm^{-3}}$ gradually strip the multi-phase ISM from the galaxy while enhancing star formation in the disk at least for 400\,Myr after the interaction with the winds. In contrast, strong ICM winds with $P_{\rm ram}/k_{\rm B}=5\times10^{5}\,{\rm K\,cm^{-3}}$, mimicking ram pressure at a cluster center \citep[e.g.,][]{jung18}, quickly remove most of the ISM, suppressing star formation on a timescale of $\sim 100\,{\rm Myr}$. However, no star formation occurs in the RPS tails due to the absence of dense molecular clouds, even when the radiative cooling rates are enhanced by the adoption of a low ICM temperature of $T=10^6\,{\rm K}$.

Although a limited number of cases have been observed thus far, RPS galaxies with massive molecular clouds in their tails are known to be gas-rich in their disks \citep{jachym17,moretti18,jachym19}. Half of these galaxies appear to be located close to a cluster center ($r_{\rm c}<300\,{\rm kpc}$), perhaps experiencing strong ram pressure. Although gas-rich galaxies are likely to be more resilient to ram pressure due to their high column density \citep{gunn72} and strong ISM pressure sustained by the active star formation~\citep[e.g.,][]{ostriker10,kimcg18}, the strong ram pressure can remove a large amount of the ISM from the galaxies. Such stripped gas may contribute to the accumulation of cold gas in the galaxy tails via radiative cooling \citep[e.g.,][]{armillotta16,armillotta17,gronke18}. The stripped ISM can also be mixed with the ICM~\citep{franchetto21}, forming H$\alpha$ and X-ray tails that are different from the pure ISM or ICM~\citep{poggianti19,sun21}. Motivated by these observational and theoretical results, we hypothesize that the abundant ISM stripped from a galaxy could facilitate molecular clump formation in the RPS tails of gas-rich galaxies. 

This study aims to understand the formation process of galaxies with multi-phase gas and young stars in their tails, called jellyfish features, via RHD simulations. Section 2 describes the RHD method and the initial conditions of our simulations. Section 3 examines the formation of molecular clouds and stars in the disks and tails of simulated galaxies and compares the findings with observed cases. Section 4 demonstrates the correlation between H$\alpha$ emission and star formation activity in an RPS tail. In Section 5, we discuss the origin of the H$\alpha$-X-ray surface brightness (SB) relation observed in the RPS tails. Finally, Section 6 summarizes the major findings.

\section{Simulations}
In this section, we describe our RHD simulations including the initial conditions of the simulated galaxies and the ICM wind, and the computation of the H$\alpha$ emissivity.

\subsection{Code}

We use \ramsesrt\ \citep{rosdahl13,rosdahl15a}, which is a RHD version of the adaptive mesh refinement code \ramses\ \citep{teyssier02}. \ramsesrt\ adopts the HLLC \citep{toro94} and the Particle-Mesh method \citep{guillet11} to solve the Euler equations and the Poisson equations, respectively. The \ramsesrt\ version used herein includes a modified photo-chemistry model for tracing the formation and destruction of molecular hydrogen \citep{kimm17,katz17} as well as the non-equilibrium chemistry and cooling of six chemical species: HI, HII, HeI, HeII, HeIII, and e$^{-}$ \citep{rosdahl13,rosdahl15a}. Atomic metal cooling at $T\gtrsim10^4$K is computed using the Cloudy cooling model~\citep{ferland98}, whereas fine structure line cooling at $T\lesssim10^4$K is computed using the cooling model of \citet{rosen95}. Radiative cooling induced by molecular hydrogen is also included~\citep{hollenbach79,halle13}. 

The spectral energy distributions of stars are obtained from the Binary Population and Spectral Synthesis model~\cite[\bpass\ version 2.0,][]{eldridge08,stanway16} based on an initial mass function (IMF) with slopes of -1.3 for stellar masses between 0.1 and $0.5\,\msun$ and -2.35 for stellar masses between 0.5 and $100\msun$ \citep{kroupa01}. We utilize a star formation model that computes star formation efficiency based on the local thermo-turbulent condition~\citep{kimm17}. Star formation is allowed in cells with hydrogen number density $n_{\rm H}$ higher than $100\,{\rm cm^{-3}}$, but most stellar particles form in cells with $n_{\rm H}>1000\,{\rm cm^{-3}}$; this is because our thermo-turbulent star formation scheme requires gravitationally bound structures~\citepalias[see][for details]{lee20}. 

The simulation box is covered with $256^3$ root cells (level 8), which are adaptively refined to resolve the local thermal Jeans length by at least eight cells until it reaches the maximum refinement level of 14. The corresponding maximum resolution is 18 pc.

Our simulations include various stellar feedback mechanisms, i.e., photoionization, radiation pressure exerted by photons at wavelengths ranging from UV to IR \citep{rosdahl13,rosdahl15a}, and Type II supernova (SN) explosions \citep{kimm15}. Additionally, the SN frequency is increased by a factor of five. Note that such a boost is necessary to reproduce the stellar mass growth and the UV luminosity functions of galaxies at $z\ge 6$ \citep[e.g.,][]{rosdahl18,garel21} or the mass fraction of stars in Milky Way-like galaxies \citep{li18}. Finally, the metal yield from SNe is neglected to allow us to distinguish the contributions from the ISM and ICM to the gas in a cell. Further details about the physical ingredients used in the simulations are presented in~\citetalias{lee20}.

\begin{figure}
\centering 
\includegraphics[width=0.45\textwidth]{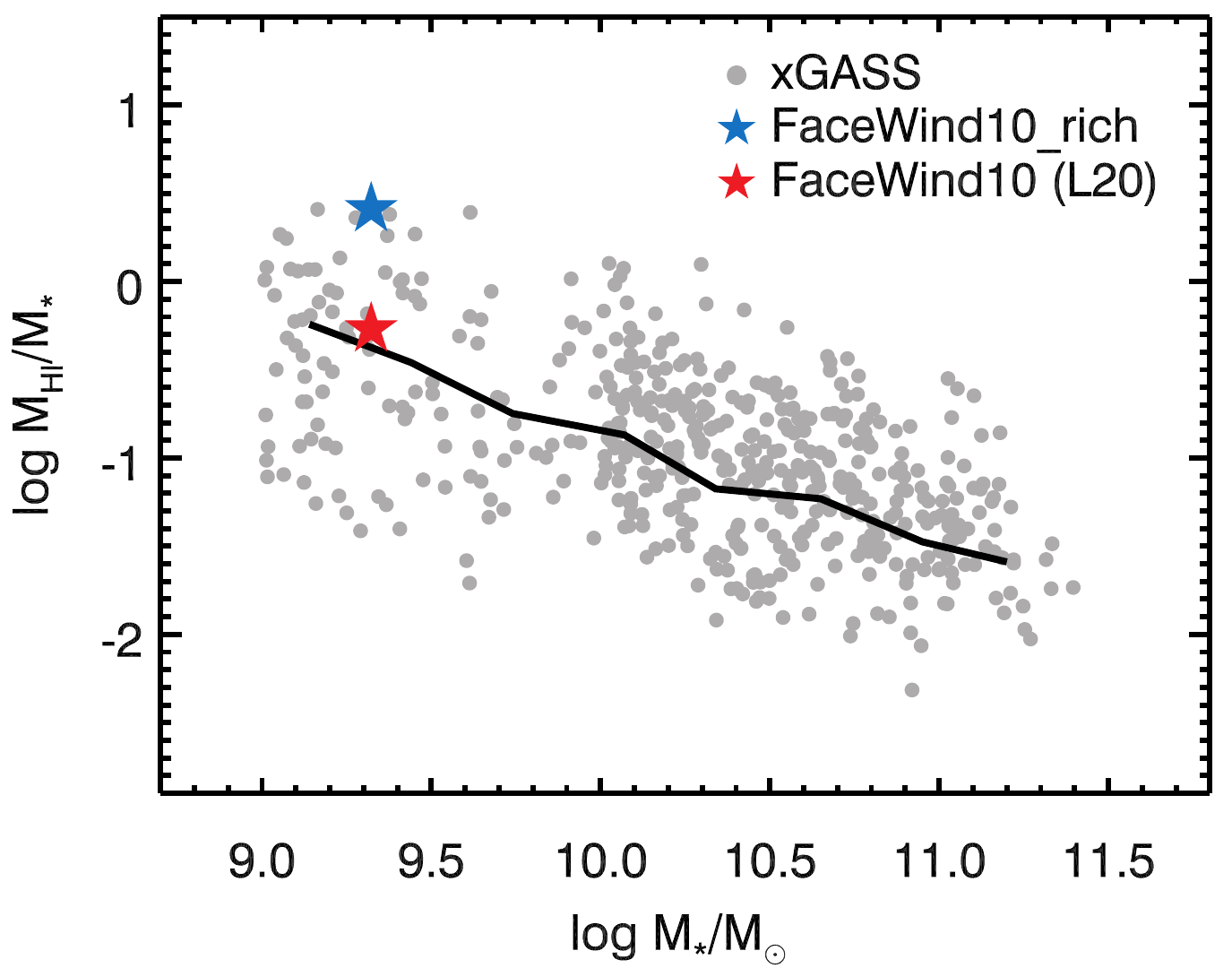}
\caption{Initial gas fraction of the simulated galaxies ( \rich; blue star, \normal; red star). Gray circles show the HI fraction-to-stellar mass relation of the local galaxies from the xGASS survey \citep{catinella18}. The black solid line denotes the average of the logarithmic HI mass fractions weighted by the local galaxy stellar mass function~\citep[see][ for further details]{catinella18}.}
\label{fig:fhi}
\end{figure}

\subsection{Initial Conditions}

To investigate the conditions for jellyfish galaxy formation, we adopt the same structural properties of the disk galaxy that were employed in \citetalias{lee20} (\normal), except that we alter the amount of disk gas mass.
The galaxies initially have a stellar mass of $2.10\times10^9 \, \msun$ with a bulge-to-total mass ratio of $f_{\rm bulge}\approx0.17$ and are embedded in a dark matter halo of mass $M_{\rm halo}=10^{11}\,{\msun}$ and virial radius $R_{\rm vir}=89\,{\rm kpc}$, as in \citet{rosdahl15b}. The disk gas metallicities are set to $Z_{\rm ISM}=0.75\,Z_{\odot}$, where the solar metallicity is $Z_{\odot}=0.0134$~\citep{asplund09}. The initial gas mass of the \normal\ galaxy \citepalias{lee20} is $M_{\rm gas}=1.75\times10^9 \, \msun$, and we adopt five times more gas mass in the \nowind\ and \rich\ runs ($M_{\rm gas}=8.75\times10^9 \, \msun$). As can be seen in  Figure~\ref{fig:fhi}, these initial gas masses represent a galaxy with a normal or high gas fraction, compared to  the local galaxies in the xGASS survey \citep{catinella18}. 

Starting from these initial conditions, the gas-rich galaxy (\rich) is evolved for $250\,\myr$ to ensure that it enters a quasi-equilibrium state before it interacts with the ICM wind. At $250\,\myr$, the mass of neutral, molecular, and ionized hydrogen, and stars in the galaxy, measured in the cylindrical volume of a radius of $r=12\,{\rm kpc}$ and a height of $z=\pm 3\,{\rm kpc}$ is $M_{\rm HI}=1.93\times10^9\,\msun$, $M_{\rm H_2}=4.90\times10^8\,\msun$, $M_{\rm HII}=3.37 \times10^8\,\msun$, and $M_{\star}=3.22 \times10^9\,\msun$, respectively. For comparison, before interaction with the ICM wind, the \normal\ galaxy has $M_{\rm HI}=6.43\times10^8\,\msun$, $M_{\rm H_2}=2.62\times10^8\,\msun$, $M_{\rm HII}=1.27 \times10^8\,\msun$, and $M_{\star}=2.18 \times10^9\,\msun$. Thus, compared to the \normal\ galaxy in \citetalias{lee20}, the \rich\ galaxy has 2.7 times more cold gas mass ($M_{\rm HI+H_2}$) or 1.8 times higher cold gas fraction ($M_{\rm HI+H_2}/M_{\star}$) before the first interaction with the ICM wind.

\begin{table}
    \centering
    \caption{Initial parameters of the simulations. From left to right, each column indicates the model name, ICM density ($n_{\rm H,\,ICM}$), ICM velocity ($v_{\rm ICM}$), ram pressure of the wind, and initial gas mass in the disk of the simulation. }
    \begin{tabular}{lcccc}
    \hline
    Model  & $n_{\rm H,\,ICM}$ &$v_{\rm ICM}$ &  $P_{\rm ram}/k_{\rm B}$ & $M_{\rm gas}$ \\ 
       & [${\cmq}$] &  [$\kms$]  & [${\rm K \,cm^{-3}}$] & [$10^9\,\msun$] \\
       \hline
     \texttt{NoWind\_rich} & $10^{-6}$ & 0 & 0 & 8.75  \\
      \texttt{FaceWind10\_rich} & $3\times10^{-3}$&$10^3$ &  $5\times 10^5$   & 8.75  \\ 
     \texttt{FaceWind10} \citepalias{lee20} & $3\times10^{-3}$ & $10^3$ & $5\times 10^5$   & 1.75 \\
    \hline
    \end{tabular}
    \label{tab:ic}
\end{table}

We impose an ICM wind from one side of the box after $150\myr$ and define this epoch as $t=0$. The ICM wind has a temperature of $T_{\rm ICM}=10^7\,{\rm K}$,  metallicity of $Z_{\rm ICM}=0.004$ $\approx 0.3\,{\rm Z_{\odot}}$, and velocity of $v_{\rm ICM}=1000~{\rm km~s^{-1}}$, based on the observations of nearby clusters \citep[e.g.,][]{tormen04,hudson10,urban17}. In our fiducial model (\rich), given its initial velocity, the ICM wind starts influencing the galaxy at $t\approx135\myr$, after the galaxy enters a quasi-equilibrium state. The wind density is set as $n_{\rm H, ICM}=3\times 10^{-3}\cmq$ to mimic the ram pressure that a satellite galaxy would experience in the central regions of clusters with $M_{200}\sim10^{14.8}\,\msun$ at $z=0$ \citep[][see their Figure 10]{jung18}. The simulations are run up to $t=366\,{\rm Myr}$ (or 516 Myr in total including the initial relaxation phase). The ICM wind of \normal\ in \citetalias{lee20} is identical to that of \rich. For comparison, we also run a control simulation without an ICM wind (\nowind).

\begin{figure}
\centering 
\includegraphics[width=0.45\textwidth]{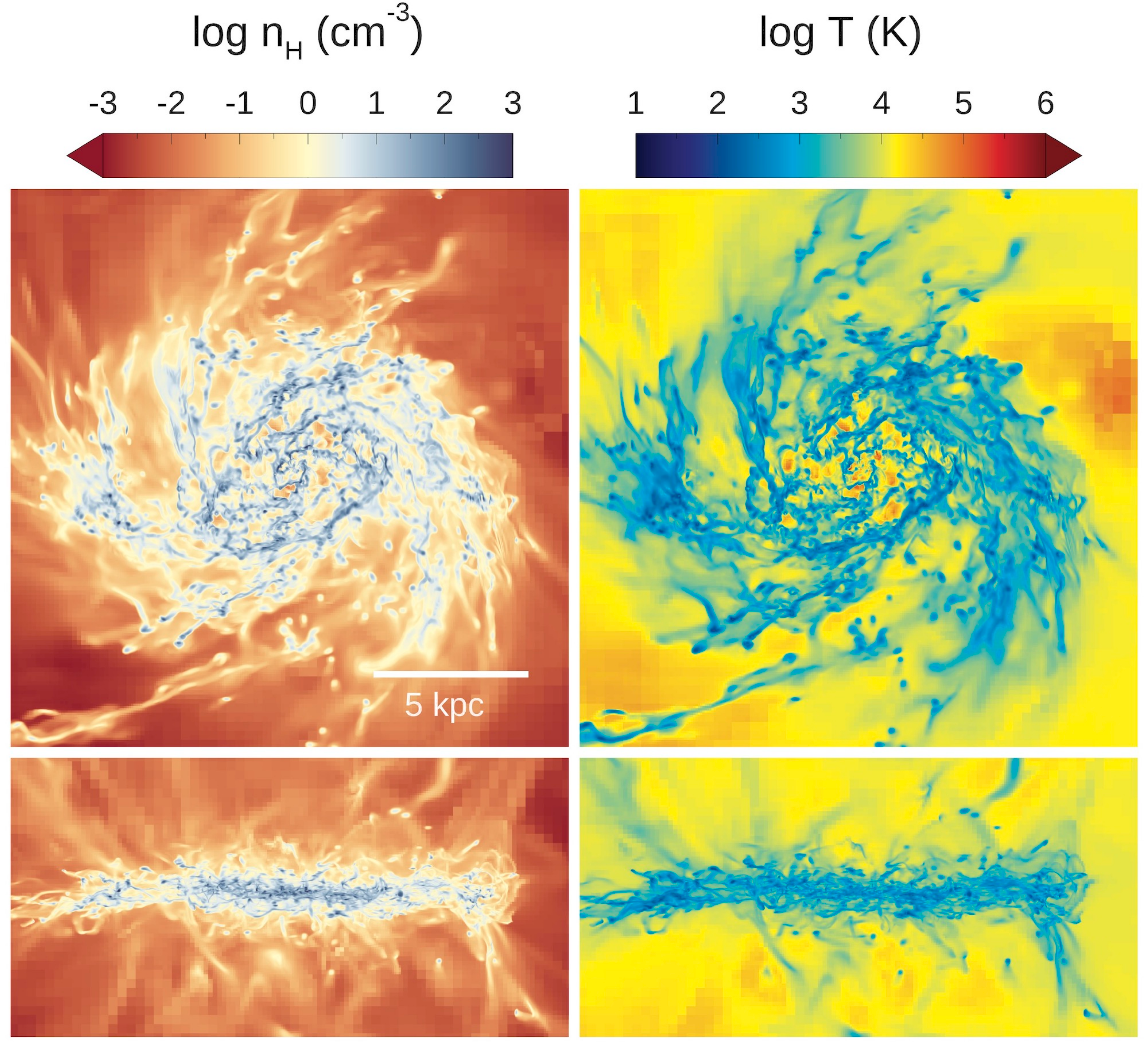}
\caption{Projected views of the gaseous disk of the simulated galaxy. The panels display the face-on (top) and edge-on (bottom) views of density-weighted distribution of the hydrogen number density (left) and temperature (right) from the \nowind\ galaxy at $t=135\,\myr$, i.e., the moment at which the ICM wind starts influencing the galaxy.}
\label{fig:reference}
\end{figure}

Figure~\ref{fig:reference} shows the projected distribution of the hydrogen number density and temperature of the galaxy at $t=135\,\myr$ in \nowind. Due to vigorous stellar feedback led by active star formation in the gas-rich disk, the gaseous disk is thicker and more feathery than that with a normal gas fraction (see Figure~1 of \citetalias{lee20}).


\subsection{Computing H$\alpha$ emission with dust}

Warm ionized gas with $T\sim10^4\,$K emits H$\alpha$ photons with $\lambda=6562.8\,{\rm \AA}$ via the recombination of ionized hydrogen with a free electron. The number of H$\alpha$ photons emitted from a cell per unit time during the recombination process is 
\begin{equation}
N_{\rm H\alpha, rec}=n_{\rm e} n_{\rm HII} \epsilon^{B}_{\rm H\alpha} (T)\alpha_B(T)\times(\Delta x)^3,
\end{equation}
where $n_{\rm e}$ and $n_{\rm HII}$ are the electron and ionized hydrogen number densities in the cell, respectively, $T$ is the temperature of the gas in the cell, $\alpha_B(T)$ is the case B recombination coefficient, $\epsilon^{B}_{H\alpha}(T)$ is the recombination fraction yielding $\rm H\alpha$ photons at $T$, and $(\Delta x)^3$ is the volume of each cell. The recombination fraction $\epsilon^{B}_{\rm H\alpha}$ is computed using the following fit to the recombination coefficients given by \citet{storey95}:
\begin{equation}
\begin{split}
\epsilon^{B}_{\rm H\alpha} (T)=8.176\times10^{-8}-7.461\times10^{-3}\log T_4 \\
+0.451 \,T_4^{-0.1013}
\end{split}
\end{equation}
where $T_4\equiv T/10^4\,{\rm K}$. The recombination coefficient $\alpha_B$ is taken from \citet{hui97}, as
\small
\begin{equation}
\alpha_B(T)=2.753\times10^{-14}{\rm cm^3\,s^{-1}}\frac{\lambda^{1.5}_{\rm HI}}{[1.0+(\lambda_{\rm HI}/2.740)^{0.407}]^{2.242}},
\end{equation}
\normalsize
where $\lambda_{\rm HI}=2\times157\,807\,{\rm K}/T$.

Another process that yields H$\alpha$ photons is collisional excitation of HI by free electrons. The number of H$\alpha$ photons emitted from a cell per unit time via collisional excitation is given by:
\begin{equation}
N_{\rm H\alpha, col}=n_{\rm e} \,n_{\rm HI} \,C_{i,j}^{\rm e} (T)\times(\Delta x)^3,
\end{equation}
where $n_{\rm HI}$ is the number density of neutral hydrogen and $C_{i,j}^{\rm e} (T)$ is the electron collisional excitation rate coefficient for a Maxwellian electron velocity distribution at $T$. Following Katz et al. ({\sl in prep.}), the electron collisional excitation rate coefficient is computed as,
\begin{equation}
C^{\rm e}_{i,j}(T)=\frac{8.628\times10^{-6}\, {\rm cm^3\,s^{-1}}}{T^{1/2}}\frac{\Upsilon_{i,j}(T)}{\omega_i}\exp\left(\frac{-E_{i,j}}{k_{\rm B}T}\right),
\end{equation}
where $\omega_i$ is the statistical weight of energy level $i$, $E_{i,j}$ is the energy difference between levels $i$ and $j$, $k_{\rm B}$ is the Boltzmann constant, and $\Upsilon_{i,j}(T)$ is the thermally averaged collision strength, which can be theoretically obtained as follows:
\begin{equation}
\Upsilon_{i,j}(T)=\int^{\infty}_{0}\Omega_{i,j}\exp\left( -\frac{E_j}{k_{\rm B}T}\right){\rm d}\left(\frac{E_j}{k_{B}T}\right),
\end{equation}
where $E_j$ is the energy of the scattered electron relative to the energy level $j$ and $\Omega_{i,j}$ is the dimension-less collisional strength that is symmetric ($\Omega_{i,j}=\Omega_{j,i}$)~\citep{dere97}. 
To compute $\Upsilon_{i,j}(T)$ for the H$\alpha$ photon emission from electron collisional excitation of HI, we use the \chianti\ database \citep[][version 10]{delzanna21}, which provides the scaled effective electron collision strength based on the rules formulated by \citet{burgess92}. We compute the collisional emission under a case B approximation for collisions up to energy level 5~\citep{dere19}.  In practice, the collisional emissivity of H$\alpha$ can be approximated as a function of temperature (Katz et al. in prep):
\begin{equation}
\epsilon_{\rm H\alpha,col}=\frac{6.01\times10^{-19}}{T^{0.230}}\exp{\left(\frac{-8.13\times10^4}{T^{0.938}}\right)}\,{\rm erg\,cm^3\,s^{-1}}.
\end{equation}
Note that we neglect the collisional H$\alpha$ emission from a cell if its net cooling timescale is less than three times the local simulation timestep to avoid spurious emission from cells where the cooling time is under-resolved\footnote{We restart a snapshot at 366 Myr from the \rich\ run by adopting a Courant number of 0.08 (an order of magnitude smaller than the typical value), and confirm that the collisional H$\alpha$ luminosity is converged by imposing the condition that there is no collisional radiation if $\tau_{\rm cool} < 3\,\Delta t_{\rm sim}$. Here we estimate the cooling time $\tau_{\rm cool}$ by dividing the thermal energy by the cooling rate in each cell.}.

The amount of dust in a cell is modelled following the prescription of \citet{laursen09}: 
\begin{equation}
n_{\rm d}=(n_{\rm HI}+f_{\rm ion}\, n_{\rm HII}) \, Z/Z_0,
\label{eq:dust}
\end{equation}
where $n_{\rm d}$ is a pseudo number density of dust grains, $f_{\rm ion}$ is the fraction of dust remaining in ionized gas, $Z$ is the gas metallicity, and $Z_0$ is the averaged metallicity of a galaxy. In this study, we adopt $f_{\rm ion}=0.01$ and $Z_0=0.01$ \citep{laursen09}. Dust attenuation is then applied by reducing the intrinsic line emission by $\exp(-\tau_{\rm d}(\lambda))$, where $\tau_{\rm d}(\lambda)$ ($=\sum_i n_{{\rm d},i}\Delta x_i\sigma_{\rm d,H}(\lambda) $) is the sum of dust optical depth along a sightline. The effective cross-section per hydrogen at wavelength $\lambda$ ($\sigma_{\rm d,H}(\lambda)$) is taken from \citet{weingartner01}, assuming Large Magellanic Cloud-type dust.

\begin{figure}
\centering 
\includegraphics[width=\linewidth]{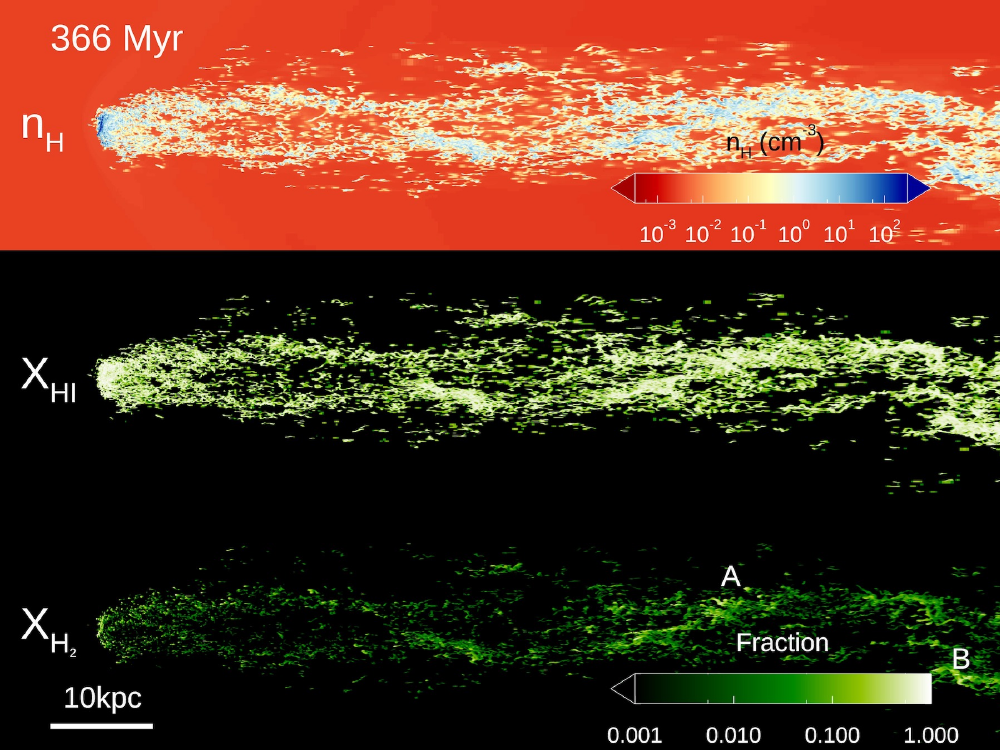}
\caption{Density-weighted projections of hydrogen number density ($n_{\rm H}$), fraction of neutral hydrogen ($X_{\rm HI}$), and fraction of molecular hydrogen ($X_{\rm H_2}$) from the \rich\ run at $t=366\,$Myr, respectively. The fractions are the mass ratios to the total hydrogen mass. Tail molecular clouds with $n_{\rm H}>100\,\cmq$ are marked by \texttt{A} and \texttt{B} in the bottom panel.}
\label{fig:gas_map}
\end{figure}

\section{Impact of Ram Pressure on Star Formation Activity}

Figure~\ref{fig:gas_map} shows the distribution of gas density, fraction of neutral hydrogen, and the fraction of molecular hydrogen in the \rich\ galaxy with prominent tail structures at the end of the simulation ($t=366\,{\rm Myr}$). In this section, we investigate how the tails develop in gas-rich galaxies after encountering strong ICM winds and how stars form in the tail regions. We also compare the cases of a gas-rich galaxy in an isolated environment (\nowind) and a RPS galaxy with a typical gas fraction (\normal\ from \citetalias{lee20}).


\begin{figure*}
\centering 
\includegraphics[width=0.825\textwidth]{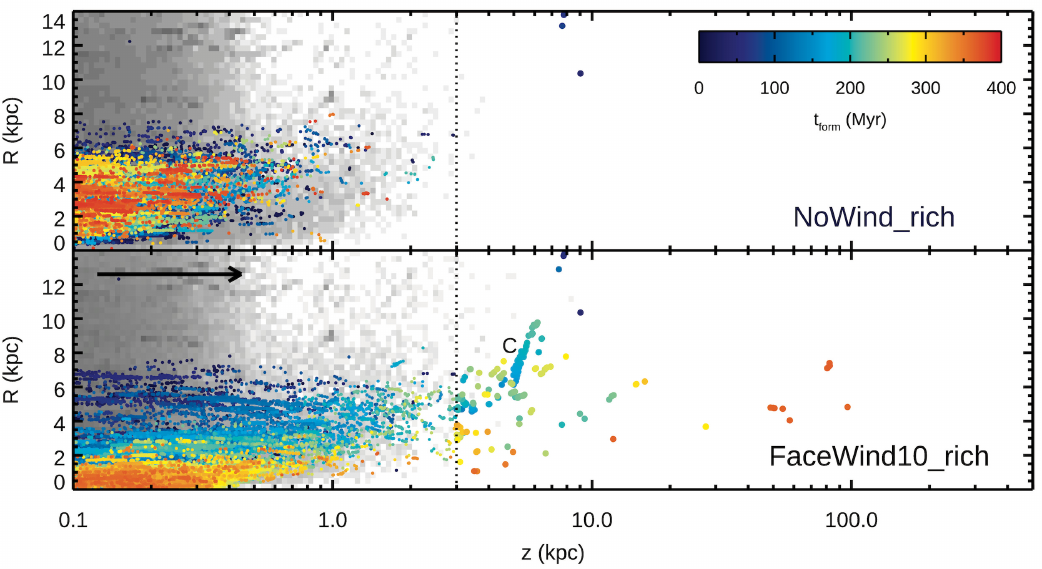}
\caption{Birthplace of stars in cylindrical coordinates in the \nowind\ (top panel) and \rich\ (bottom panel) runs. The coordinate systems are defined from the center of stellar mass and the galactic mid-plane is chosen in the $XY$ plane. The black arrow in the bottom panel indicates the ICM wind direction. The color code denotes the birth epoch of each stellar particle formed after the ICM wind is launched. The gray shades depict the distribution of stellar particles formed before $t=0$. The vertical dotted line indicates the height ($z=3\,$kpc) above which no stars form after the wind starts influencing the galaxy ($t>135\,\myr$) in the \nowind\ run. We thus adopt $z=3\,$kpc as the separation between the disk and tail. A significant star formation event in the tail region is marked as \texttt{C}. }
\label{fig:sf_region}
\end{figure*}

\begin{figure}
\centering 
\includegraphics[width=\linewidth]{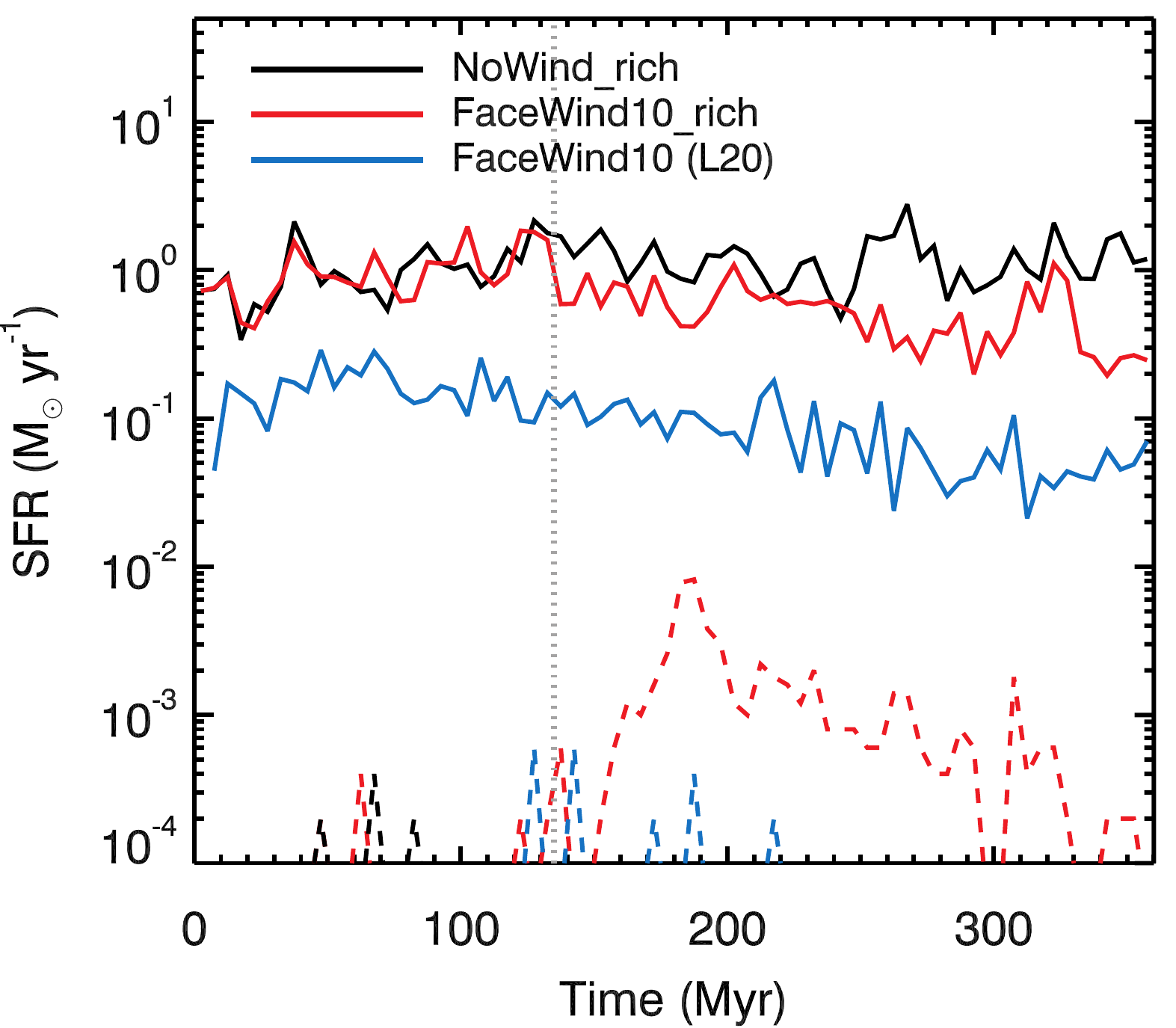}
\caption{Star formation rate evolution in the galactic disk (solid lines) and in the tail (dashed lines). The three different colors represent the \texttt{NoWind\_rich} (black), \texttt{FaceWind10\_rich} (red), and \texttt{FaceWind10} (blue, \citetalias{lee20}) runs. The gray vertical dotted line at $t=135\,$Myr denotes the epoch at which the wind front reaches the galaxy. In the \rich\ run, the star formation rates in the galactic disk decrease once the ICM wind encounters the disk, whereas the star formation in the tail commences and is most noticeable $\approx 50\,{\rm Myr}$ from the start of the ICM--ISM interaction  (i.e., $t\approx185\,$Myr).}
\label{fig:sfr}
\end{figure}

\subsection{Star formation in the Disk}

Figure~\ref{fig:sf_region} shows the birthplace of individual stellar particles in the \nowind\ (top panel) and \rich\ (bottom panel) runs at $z>0.1\,$kpc between $0\le t \le 366\,{\rm Myr}$, where $z$ is the vertical distance from the mid-plane of the disk. The galactic mid-plane is defined from the center of galaxy stellar mass in the $XY$ plane and the radial distance ($R$) is measured using the cylindrical coordinate system . For comparison, the stars born before the wind is launched (i.e., $t< 0\,{\rm Myr}$) are shown in gray. In \nowind, no stellar particles form at $z>3\,{\rm kpc}$ after $t=135\myr$, corresponding to the epoch at which the strong ICM winds start influencing the \rich\ galaxy. Thus, we define the vertical distance of $3\,{\rm kpc}$ as the separation between the tail and disk stellar populations. Notably, this scale is comparable to the height of the cylindrical volume enclosing more than $95\%$ of the total cold gas (HI+H$_2$) in the gas-rich galaxy.

We find that star formation in the disk is significantly suppressed under strong ram pressure. In Figure~\ref{fig:sf_region}, while no particular trend is seen in the radial distribution of new stars in the \nowind\ galaxy, the star-forming region shrinks over time in the \rich\ galaxy because of the stripping of the gaseous disk. We estimate the truncation radius ($r_t$) using the Gunn-Gott criterion~\citep{gunn72} by balancing the gravitational restoring force and ram pressure as, $\rho_{\rm ICM}v^2_{\rm ICM} = -\Sigma (r_c) \partial \Phi (r_c,z)/\partial z$, where $\Phi (r_c,z)$ is the gravitational potential obtained from the sum of the matter components (gas, stars, and dark matter) at a radius $r_c$ and a vertical height $z$ in the cylindrical coordinate system, and $\Sigma (r_c)$ is a gas column density at $r_c$. The column density is measured from the gas component in the concentric shell with a radius $[r_c,r_c+\Delta r]$ and a height $|z|<3\,{\rm kpc}$, following the definition of the galactic disk in this study. Next, we compute the gravitational restoring force and ram pressure at a disk thickness of $H=\sqrt{\int\rho z^2{\rm d}V/\int\rho {\rm d}V}$, where $\rho$ is the total matter density (gas, stars, and dark matter) and $dV$ is the cylindrical volume element inside the scale-length ($l$) of the cold gas (HI+H$_2$) disk. The disk thickness, scale-length, and truncation radius just after the galaxy encounters the ICM wind ($t\approx140\myr$) are $H=0.416\,{\rm kpc}$, $l=2.56\,{\rm kpc}$ and $r_t=2.74\,{\rm kpc}$, respectively. Indeed, we confirm that the gaseous disk at $r>r_{\rm t}$ is largely ($\sim90\%$) stripped within 125 Myr. Accordingly, almost all stars ($99.4\%$) form inside the truncation radius at $t>250\,{\rm Myr}$ and some stellar particles form outside the disk when the disk is first perturbed by the ICM wind ($t\sim150$--$250\,{\rm Myr}$).

\begin{figure}
\centering 
\includegraphics[width=0.475\textwidth]{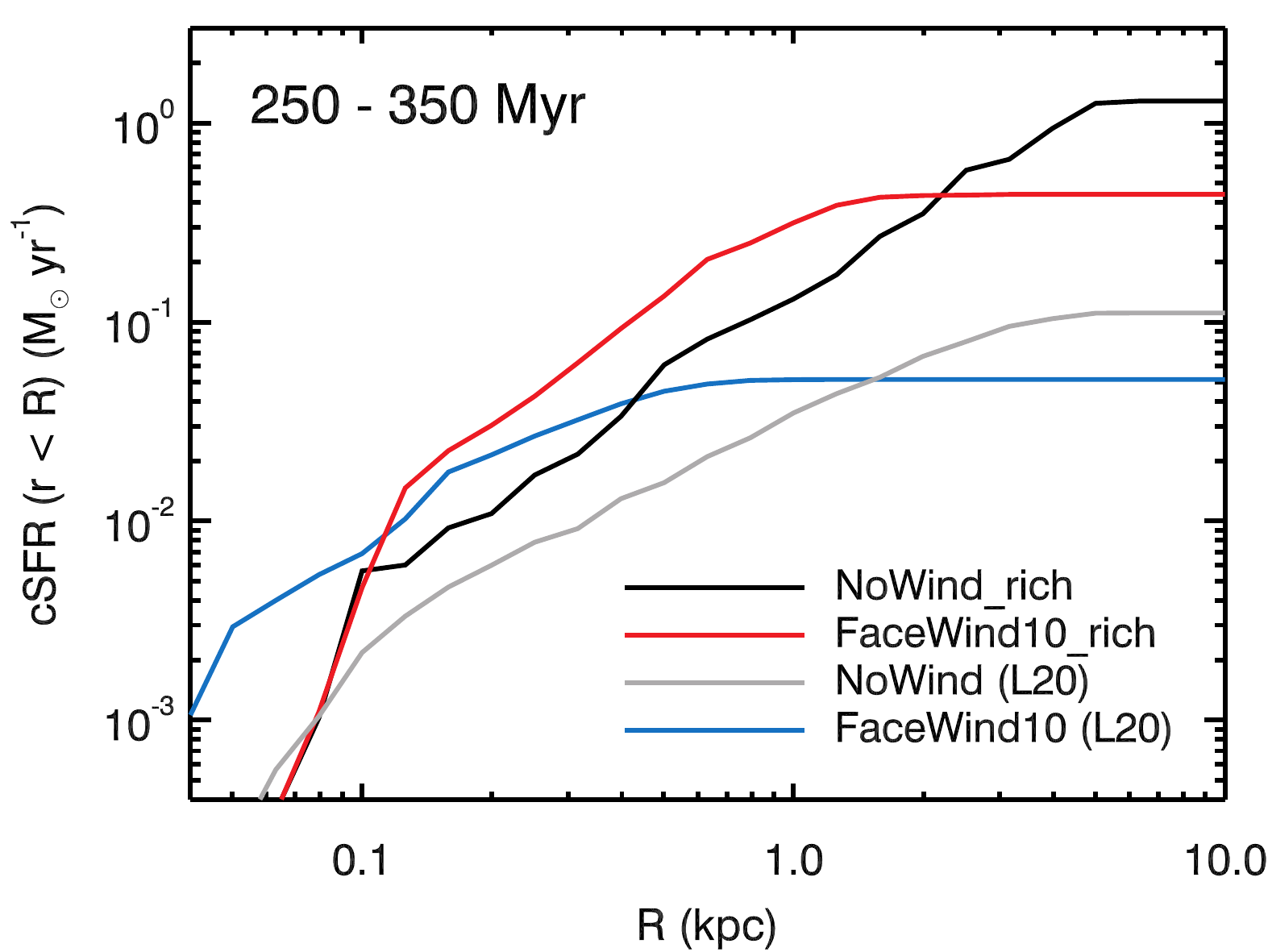}
\caption{Cumulative star formation rates (cSFRs) as a function of the cylindrical radius of the disk ($R$), averaged over $250\le t \le 350$\,Myr. For comparison with \normal, we also plot the cSFR of the \texttt{NoWind} galaxy of \citetalias{lee20} (grey solid line). The galaxies interacting with the ICM wind (red and blue lines) show enhanced star formation in the central region ($R\la 1\,{\rm kpc}$), but it is suppressed at $R\ga 1\,{\rm kpc}$, compared to their counterparts without winds (black and grey lines).}
\label{fig:csfr}
\end{figure}

Figure~\ref{fig:sfr} compares the star formation rates in the disk and tail of the three simulations. Once the strong ICM wind begins to influence the simulated galaxies, the star formation in the disk decreases by $\approx 40\%$ every $100\,{\rm Myr}$ in the \rich\ and \normal\ galaxies. During $135<t<366\,{\rm Myr}$, the stellar mass in the \rich\ galaxy increases by $1.22\times10^8\,\msun$, which is less than half the stellar mass formed in the \nowind\ galaxy ($\Delta M_{\star}=3.01\times10^8\,\msun$) during the same period. The star formation activity is suppressed from outside to inside due to disk truncation. In contrast, as illustrated in Figure~\ref{fig:csfr}, star formation is enhanced in the central region ($r<1\,$kpc) by a factor of $\approx 2$ due to the compression of the ISM. The same trend is observed in the galaxy with a normal gas fraction. In contrast, the disk in the \nowind\ galaxy maintains its star formation rate profile during the entire time period. This demonstrates that the strong ram pressure efficiently quenches the star formation not only in galaxies with a typical gas fraction but also in a gas-rich galaxy by stripping away the gaseous disk in extreme environments, such as the central region of galaxy clusters.

\begin{figure}
\centering 
\includegraphics[width=0.475\textwidth]{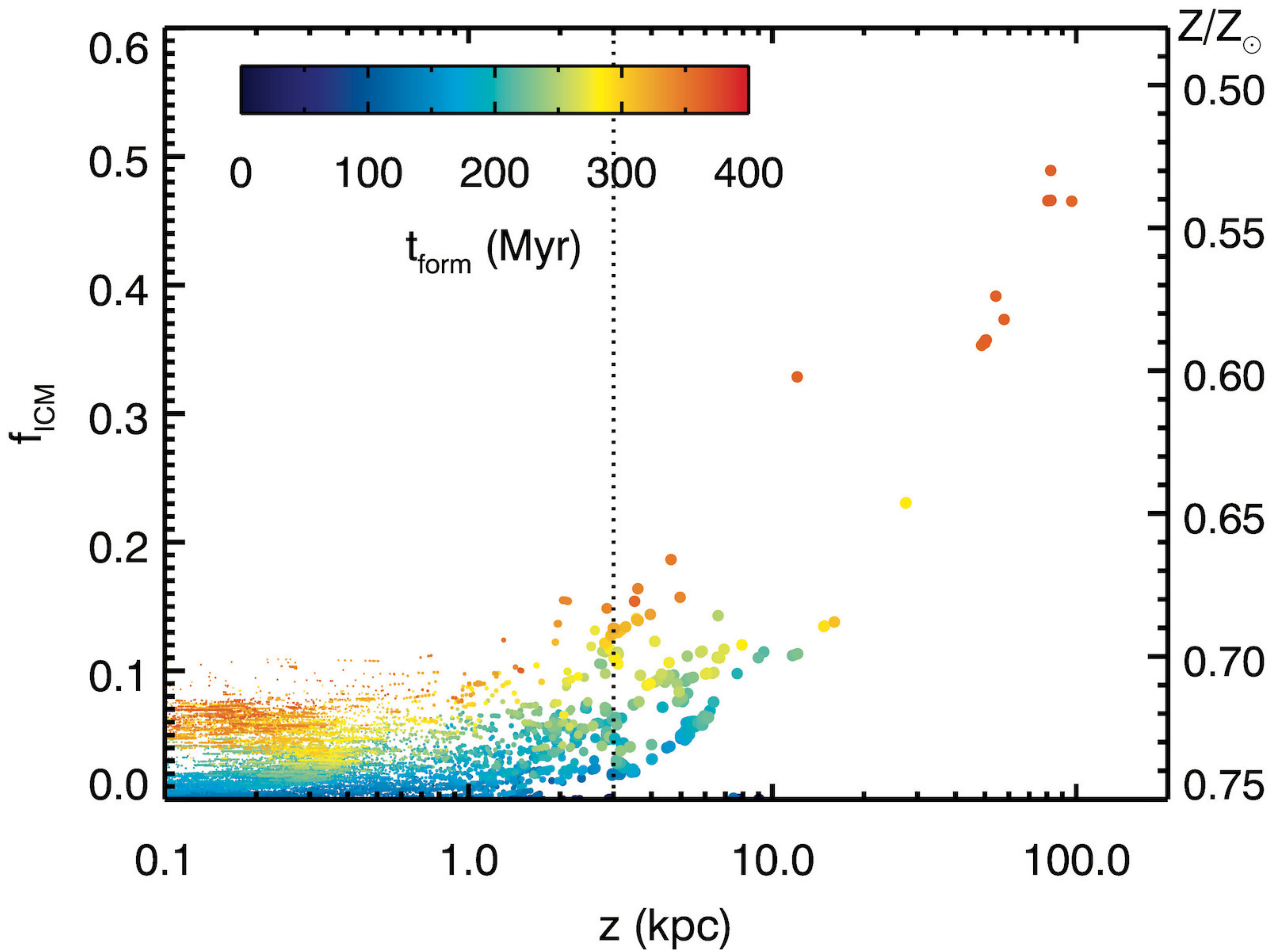}
\caption{Fraction of the ICM-origin gas ($f_{\rm ICM}$) in each stellar particle as a function of vertical distance\,($z$). $f_{\rm ICM}$ is estimated from the metallicity of stellar particles\,($Z$) by differentiating the contribution from the ISM ($Z_{\rm ISM}=0.75\,Z_\odot$) and ICM ($Z_{\rm ICM}=0.3\,Z_\odot$). The color code denotes the formation epochs of stellar particles, and the vertical dotted line indicates the separation between the tail and disk. The increasing trend of $f_{\rm ICM}$ indicates that more distant stars form in the clouds which have mixed more efficiently with the cooled ICM.}
\label{fig:mixing}
\end{figure}

\subsection{Star formation in the Tail}
Figure~\ref{fig:sf_region} shows that stars form in the stripped wake of the gas-rich galaxy at $z>3\,{\rm kpc}$. By the end of the simulation, a total stellar mass of $3.16\times10^5\,\msun$ is produced in the tail ($z>3\,$kpc), with $\sim90\%$ of it located in $3<z<10\,{\rm kpc}$. The bluish trail marked by \texttt{C} ($R\approx6-10\,$kpc and $z\approx5-6\,$kpc) in the bottom panel of Figure~\ref{fig:sf_region} is particularly noteworthy. The formation time of each stellar particle is tightly correlated with the distance from the galactic center, indicating that the stellar particles are formed inside clouds moving outward. We visually confirm from the tail region close to the galaxy that the ISM gas is first stripped away from the disk and then collapses to form giant clouds at $z\sim5\,$kpc, which in turn produce two adjacent stellar clumps of mass of $M_{\star}=3.0\times10^4\,\msun$ and $1.3\times10^4\,\msun$.

The gas-rich galaxy exhibits a remarkably higher star formation rate in its stripped wake than the \normal\ galaxy (dashed lines in Figure~\ref{fig:sfr}). Its star formation rate is still lower than the galactic star formation rates, but it increases to $0.001$--$0.01\,\msunyr$ for $\sim 50\,\myr$ once the disk encounters the wind. The star formation rate in the tail is comparable to that of D100 (${\rm d}M_{\star}/{\rm d}t=3.9\times10^{-3}\,\msunyr$), a spiral galaxy with $M_{\star}\sim2\times10^9\,\msun$ experiencing strong ram pressure stripping in the Coma cluster~\citep{jachym17}, albeit with some differences (see Section~\ref{sec:com_obs}). In contrast, no significant star formation occurs in the stripped wake of the \normal\ galaxy.

Figure~\ref{fig:mixing} hints at different origins for the stars formed in the stripped wake and the disk. We remind the readers that the fraction of gas cooled from the ICM can be directly inferred from the stellar metallicity $Z_\star$, since the initial metallicities of the ISM and ICM are fixed ($Z_{\rm ISM}=0.75\, Z_{\odot}$ and $Z_{\rm ICM}=0.3\, Z_{\odot}$, respectively) and the metal enrichment due to SNe is turned off. The ICM-origin fraction of a gas cell or a stellar particle with a metallicity $Z$ is computed as $f_{\rm ICM}=1-f_{\rm ISM}=(Z_{\rm ISM}-Z)/(Z_{\rm ISM}-Z_{\rm ICM})$. First, more than 97.5\% of the new stars in the disk ($z<3\,$kpc) show $f_{\rm ICM}$ lower than 7.6\%. At a fixed vertical distance, stars formed at a later epoch have slightly higher $f_{\rm ICM}$, indicating that a fraction of the ICM wind continuously accretes on to the central disk. 
Indeed, the column density of the ICM that encounters the disk during the interaction ($n_{\rm H,ICM}\,v_{\rm ICM}\Delta t$) amounts to a few percent of the typical $N_H$ in the central disk or roughly ten percent of that in the outer gaseous disk ($r\approx 3\,R_{\rm 1/2}$). 
However, the increase of $f_{\rm ICM}$ is not dramatic ($< 0.1$) during $t\sim150-366$\,Myr, indicating that a dense ISM is largely shielded from strong ICM winds. 
This is likely due to the strong turbulent pressure afforded by the interplay between gravity and stellar feedback, as demonstrated by \citetalias{lee20}.
Second, new stars formed at $z=3-10\,$kpc also primarily originate from the ISM. The ICM fraction that contributes to new star particle formation in the tail is less than 20\% ($f_{\rm ICM}<0.2$). On the other hand, the new stars located far behind the galactic plane ($z>10\,$kpc) form in clouds well mixed with the ICM and only half of the stellar mass originates from the stripped ISM. The mass of stars formed in the distant tail is considerably lower ($M_{\star}=2.40\times10^4\,\msun$) than that formed near the galactic plane, but their presence implies that molecular clouds can form in the RPS tail. In contrast, dense clouds with $n_{\rm H}>100\,{\rm cm^{-3}}$ do not form in the stripped wake of the \normal\ galaxy at $t>200\myr$.

\begin{figure}
\centering 
\includegraphics[width=\linewidth]{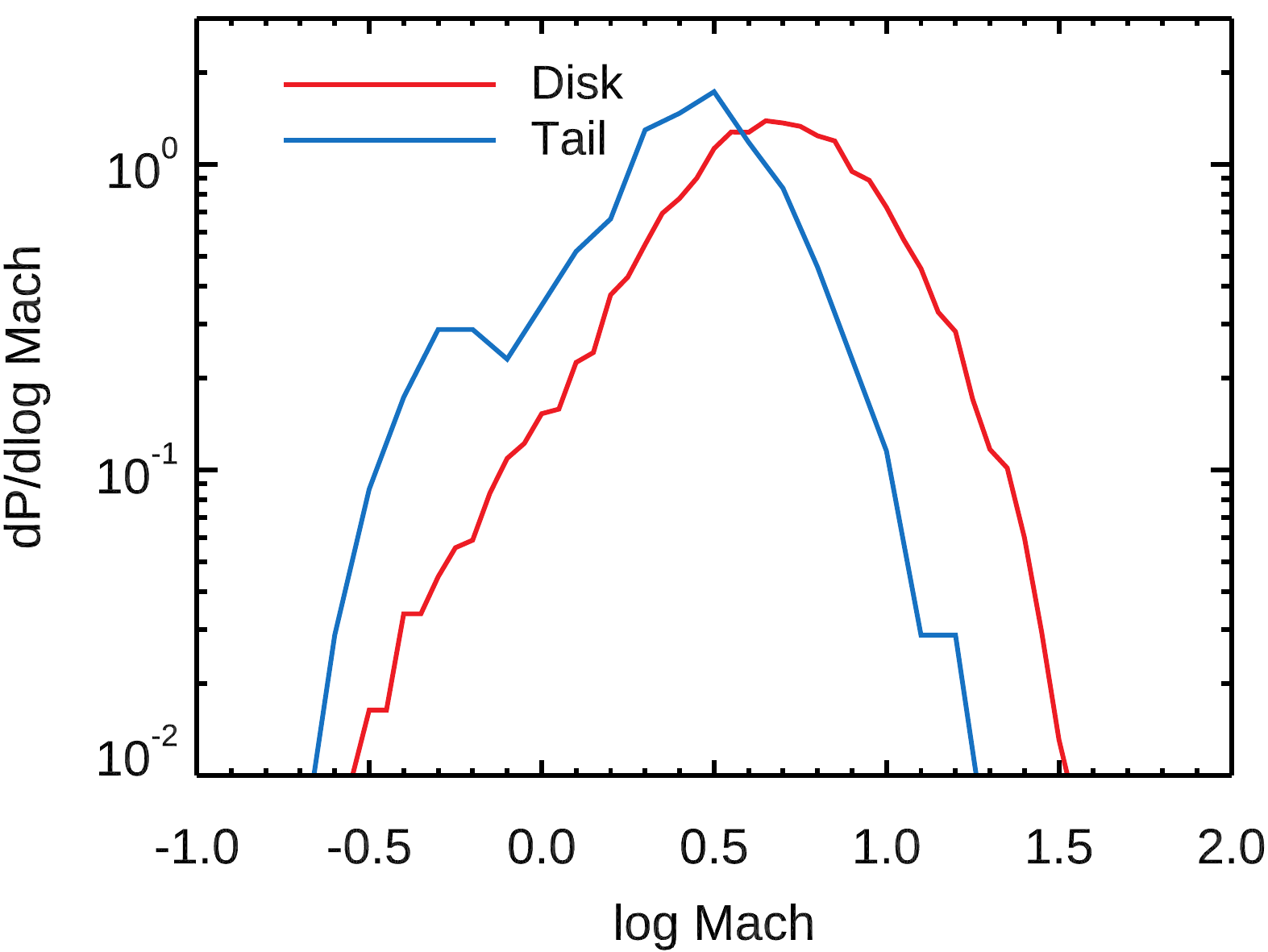}
\caption{Turbulent properties of star-forming candidate cells with $\nH>100\,\cmq$ in the disk (red) and tail region (blue). The turbulent Mach number is estimated from the six immediate neighboring cells (see text). Note that star-forming regions in the RPS tail are not more turbulent than the galactic disk.}
\label{fig:mach}
\end{figure}

We then examine the dynamical state of tail gas clouds in the \rich\ galaxy by measuring the turbulent Mach number. The turbulent Mach number is computed as $\mathcal{M}=\sigma_{\rm gas}/c_s$, where $\sigma_{\rm gas}=|\partial \vec{v}/\partial \vec{x}|\Delta x$ is the turbulent velocity measured using the six immediate neighboring cells, and $c_s$ $=\sqrt{\gamma P_{\rm th}/\rho_{\rm gas}}$  is the local sound speed with an adiabatic index of $\gamma=5/3$, where $\rho_{\rm gas}$ and $P_{\rm th}$ are the gas density and thermal pressure of the cell of interest, respectively. Figure~\ref{fig:mach} shows the probability density function of the Mach number distribution (${\rm d}P/{\rm d}\log\,\mathcal{M}$) of gas cells with $n_{\rm H}>100\,{\rm cm^{-3}}$ in all the snapshots at $t>135\,$Myr, after which the ICM wind influences the gas-rich galaxy. The majority of the dense cells in the disk and tail have Mach numbers higher than 1, indicating the presence of supersonic turbulence. Unlike the naive expectation that the dense gas in the tail may be significantly perturbed by the ICM wind, the dense gas with $n_{\rm H}>100\,{\rm cm^{-3}}$ in the tail region  is in fact slightly less turbulent than that of the disk. Furthermore, the dense gas in the disk and tail exhibit virial parameter\footnote{The virial parameter for individual cells is computed as $\alpha_{\rm vir}\equiv 2|E_{\rm kin}|/E_{\rm pot}\approx 5(\sigma_{\rm gas}^2+c_s^2)/\pi \rho_{\rm gas} G \Delta x^2$.} distributions similar to each other. These results demonstrate that the tail has a quiescent star formation activity not because its dense clouds are more turbulent than those in the galactic disk, but simply because the gas reservoir for star formation is limited in the tail (e.g., $M_{\rm dense,disk}=1.73\times10^{8} \,\msun$ vs. $M_{\rm dense,tail}=2.16\times10^{6}\,\msun$ for gas with $n_{\rm H}>100\,{\rm cm^{-3}}$ at $t=250\,{\rm Myr}$).

\begin{figure}
\centering 
\includegraphics[width=\linewidth]{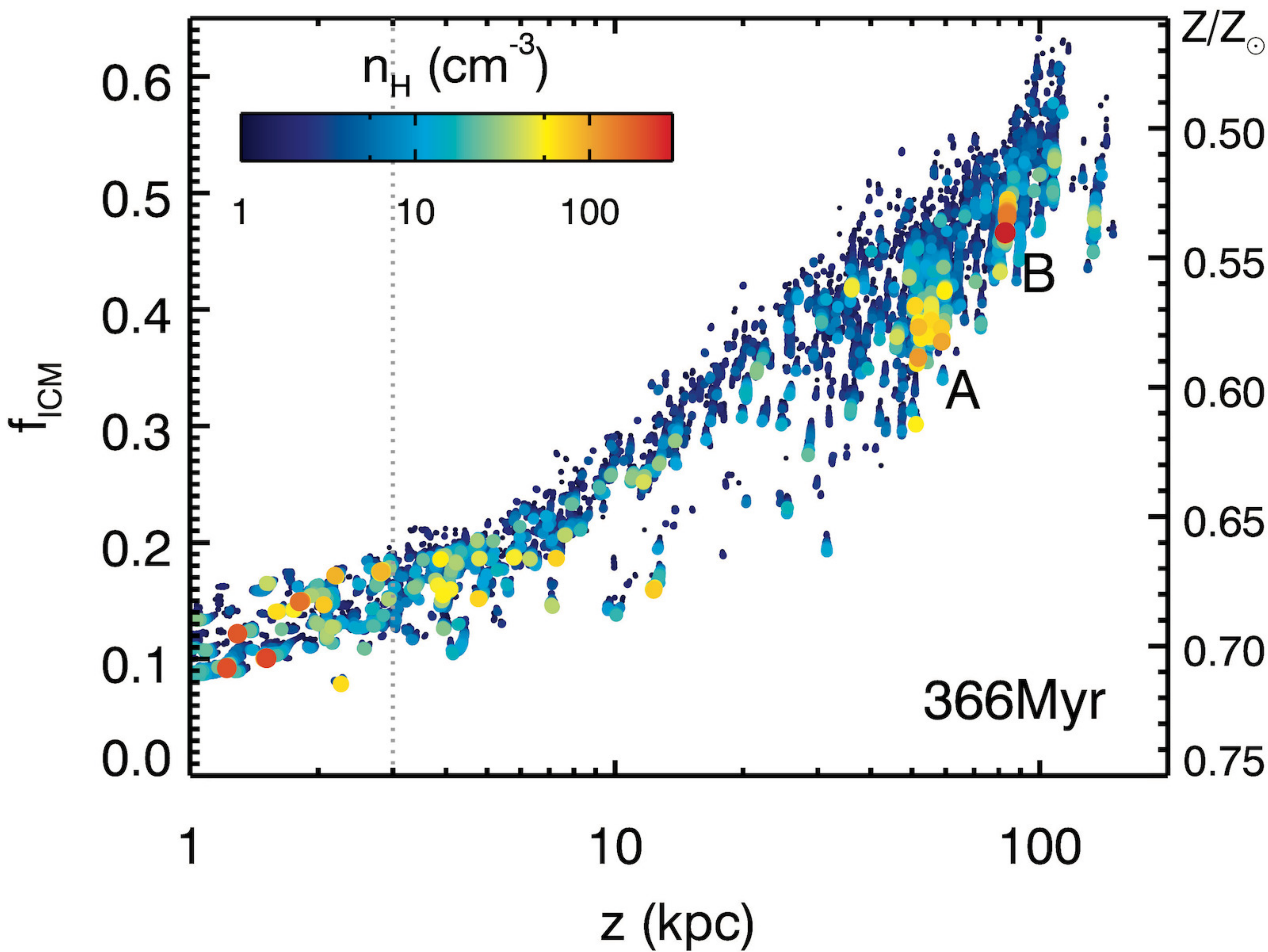}
\caption{Fraction of the molecular hydrogen stemming from the ICM as a function of vertical distance at $t=366\,$Myr. The color code denotes the hydrogen number density of the molecular gas in a cell. The symbol size indicates the mass of the molecular gas ranging from $150\,\msun$ to $2.7\times10^4\,\msun$. Only the gas cells with $\nH>1\,\cmq$ are displayed. Two prominent molecular clouds with $\nH>100\,\cmq$ in the tail region ($z>3\,$kpc, demarcated by the vertical dotted line) are marked by \texttt{A} and \texttt{B}, which are also presented in the bottom panel of Figure~\ref{fig:gas_map}.}
\label{fig:h2_mixing}
\end{figure}

\begin{figure}
\centering 
\includegraphics[width=\linewidth]{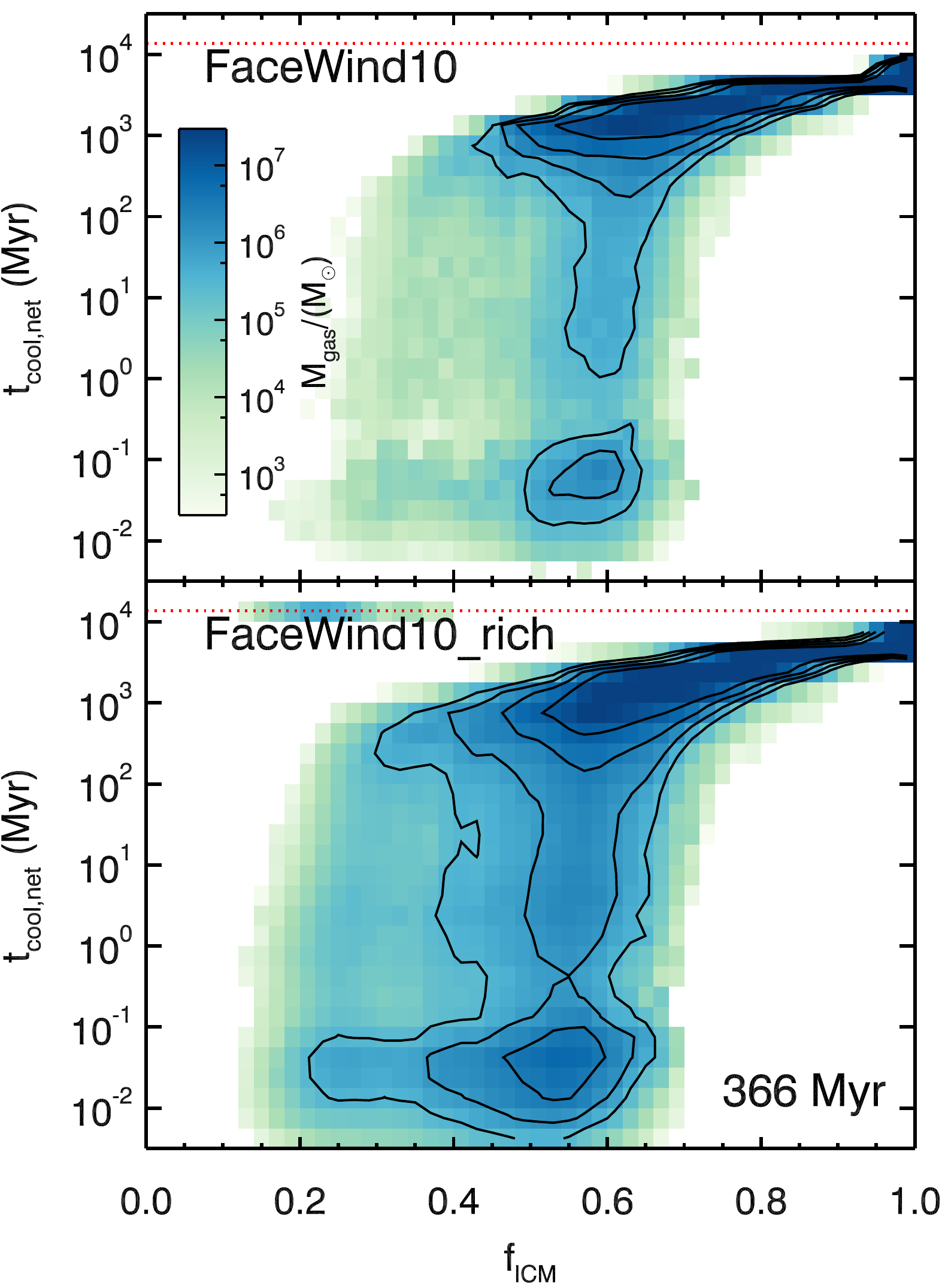}
\caption{Net cooling time of the gas in the RPS tail as a function of $f_{\rm ICM}$ at $t=366\myr$. The top panel shows the distribution for \normal, whereas the \rich\ case is shown in the bottom. We measure $t_{\rm cool,net}$ for the cells with HII mass fractions higher than 0.5 to focus on potentially cooling regions. The net cooling time is computed by considering the cooling and heating due to local radiation.  Gas cells being heated or cells with the cooling time greater than the Hubble time (red dotted lines), i.e. $t_{\rm cool,net}>13.7\,{\rm  Gyr}$, are all indicated as $t_{\rm cool,net}=13.7\,{\rm  Gyr}$. The color code indicates the total gas mass in each bin. }
\label{fig:tcool}
\end{figure}

\subsection{Origin of the Star-Forming Gas in the Tail}

In the \rich\ galaxy, the cold dense ISM is disrupted by the interaction with the ICM. However, an RPS tail may provide a more favorable environment for gas cooling and molecular gas formation than the galactic disk vicinity, as the tail contains few young stars that emit ionizing photons. Furthermore, as hinted in Figure~\ref{fig:mixing}, the mixing with the stripped ISM in the tail increases the density and metallicity of the ICM, leading to enhanced gas cooling. In such environments, the formation of dust and molecular hydrogen is likely to become efficient~\citepalias[e.g., see Sec. 2.1.3 of][and references therein]{lee20}.

As shown in Figure~\ref{fig:gas_map}, the stripped wake clearly comprises molecular clouds which sometimes have $n_{\rm H}>100\,\cmq$ (marked as \texttt{A} and \texttt{B} in the bottom panel).  Figure~\ref{fig:h2_mixing} further shows that more distant cells contain more molecular hydrogen originating from the ICM. In the vicinity of the disk, the dominant source of the molecular hydrogen ($\sim90\%$) is the ISM and the rest is that coming from the ICM due to strong ram pressure. However, once the stripped tail is pushed out to 100 kpc from the galaxy, approximately half of the total molecular gas is cooled from the ICM,  consistent with recent observational and numerical findings \citep{franchetto21,tonnesen21}. Two dense regions at $z\approx 60\,{\rm kpc}$ (marked as \texttt{A}) and $z\approx 80\,{\rm kpc}$ (marked as \texttt{B}) are particularly interesting as they develop near the end of the simulation ($t=366\,{\rm Myr}$). Specifically, a portion of the stripped ISM is spread over a large volume of $\sim(10\,{\rm kpc})^3$, and this partially ionized, intermediate-density ($\nH\sim1\,\cmq$) and porous medium collapses to form dense clumps.

To understand the formation of molecular hydrogen in the RPS tail, we compute the net cooling timescale $t_{\rm cool,net}\equiv E_{\rm int} / \mathcal{C}_{\rm net}$, where $E_{\rm int}$ is the internal energy. The net cooling rate ($\mathcal{C}_{\rm net}$) is computed by considering the heating due to local radiation fields and cooling due to atomic and metallic species, as in the simulation. Figure~\ref{fig:tcool} illustrates the net cooling time of the gas in the RPS tails in \normal\ and \rich. In the tail region of the \normal\ galaxy, the typical density and temperature of the ionized ($f_{\rm HII}>0.5$) gas, which can potentially cool and contribute to the total HI, are low ($\nH\approx 0.01\,\cmq$) and hot ($T\approx 9\times10^6\,{\rm K}$), respectively. The median net cooling time is $t_{\rm cool,net}\approx 600\,{\rm Myr}$, which is twice the simulation duration. A portion of the gas with  $f_{\rm ICM}\sim0.5-0.7$ has a short cooling time of $t_{\rm cool,net}<1\,$Myr, but its total mass is not significant. This explains why few dense molecular clouds form in the tail of the galaxy with a normal gas fraction. 

In contrast, in the \rich\ galaxy, the amount of the ionized ISM-origin gas in the tail is approximately 20 times larger than that in the \normal\ galaxy. The presence of abundant warm ionized gas results in a substantial peak at $t_{\rm cool,net}<1\,\,{\rm Myr}$, making the $t_{\rm cool,net}$ distribution clearly bimodal. The median density and temperature of these cells in the tail are also higher ($\nH\approx 0.1\,\cmq$) and cooler  ($T\approx4\times10^4\,$K), respectively, and the cooling times are significantly shorter than those in \normal. In such conditions, the gas freely collapses within $100$--$200\,{\rm Myr}$. For comparison, pure ICM gas cannot cool or collapse within a Gyr. Thus, our numerical experiments support the observational interpretation that molecular clouds form {\it in-situ} in the distant RPS tails~\citep{jachym17,moretti18,jachym19} by increasing the tail gas density because of the mixing and enhanced cooling due to the stripped ISM.

\subsection{Comparisons with Observations}
\label{sec:com_obs}

Several observations have measured the amount and distribution of molecular gas in the wakes of RPS galaxies using CO emission lines. \citet{verdugo15} detected molecular gas amounting to $\sim10^6\,\msun$ in the RPS tail of NGC 4388 in the Virgo cluster. In contrast, ESO 137-001 possesses a large amount of molecular hydrogen of mass $\sim10^9\,\msun$ in its tail~\citep{jachym14} and a similar amount of molecular hydrogen is present in the tail of D100 in the Coma cluster~\citep{jachym17}. Furthermore, \citet{moretti18} found $M_{\rm H_2}\sim10^9\,\msun$ of molecular hydrogen in the tails of four massive jellyfish galaxies with stellar mass of $\sim3\times10^{10-11}\,\msun$. ESO 137-002 is one more case recently reported to have abundant molecular gas ($\sim5.5\times10^9\,\msun$) in its disk and tail~\citep{laudari21}. The amount of molecular hydrogen detected in the tail of ESO 137-002 is $M_{\rm H_2}\sim2.2\times10^8\,\msun$. These galaxies with massive molecular tails are observed to either currently have gas-rich disks or to have possessed them until recently~\citep{jachym14,jachym17,moretti18,jachym19,laudari21}; thus, the amount of gas in the infalling galaxies is likely a key property for forming jellyfish features, which is consistent with our experiment wherein only the \rich\ galaxy develops the prominent RPS tails.

\begin{figure*}
\centering 
\includegraphics[width=1\textwidth]{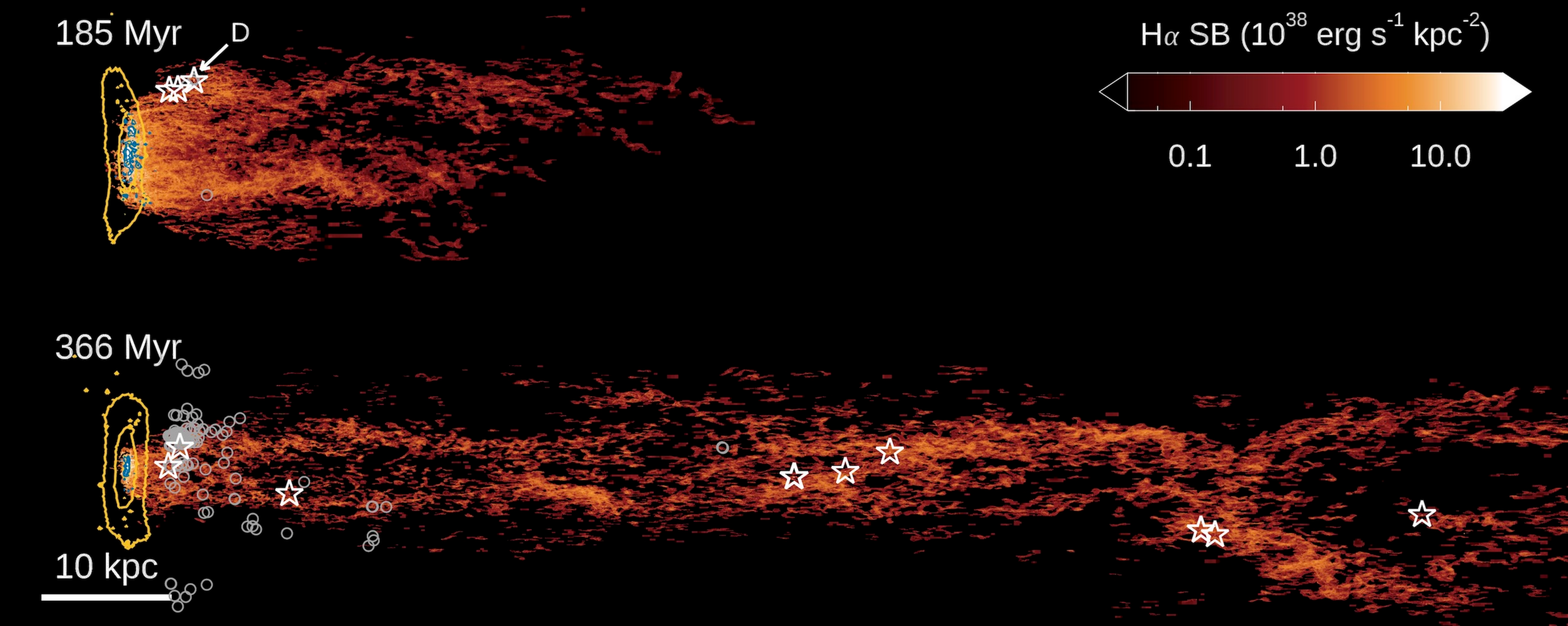}
\caption{Dust-obscured H$\alpha$ map of the gas-rich galaxy (\rich) at $t=185$ (top) and 366\,Myr (bottom). Yellow and blue contours show the distribution of all stars and the stars younger than 20\,Myr in the galactic disk, respectively. White star symbols mark the locations of stellar particles younger than 20\,Myr in the RPS tail ($z>3\,$kpc). Older stellar particles formed in the tail at $100 < t_{\rm form} < 344\,\myr$ are denoted as open gray circles. A clump of young stars generating bright H$\alpha$ cores ($F_{\rm H\alpha}>10^{41}\,{\rm erg\,s^{-1}\,kpc^{-2}}$) in the stripped wake is marked as \texttt{D} in the upper panel.}
\label{fig:halpha}
\end{figure*}

Of those, D100 in the Coma cluster is probably the most comparable example to the \rich\ galaxy. D100 has a stellar mass of $2.1\times10^9\,\msun$ \citep{yagi10}, which is similar to that of the \rich\ galaxy ($M_{\star}=3.3\times10^9\,\msun$). Since the orbital velocity of D100 is $v\sim3000-4000\,{\rm km\,s^{-1}}$ and the ICM density is $\sim3.3\times10^{-27}\,{\rm g\,cm^{-3}}$ at the projected distance of 240\,kpc \citep{jachym17}, the ram pressure currently exerted on D100 is also very strong (6--10 times larger than that of \rich). Moreover, since the tail of D100 is nearly perpendicular to the direction to the cluster center, D100 is likely passing through its pericenter, suggesting that it has been exposed to strong ram pressure for $\sim 200\,{\rm Myr}$\footnote{We infer the orbital motion of D100 assuming a dark matter halo of mass  $M_{\rm 200}=2.7\times10^{15}\,\msun$ and a radius of $R_{200}=2.9\,$Mpc, estimated for the Coma cluster \citep{kubo07}. We adopt the Navarro-Frenk-White profile \citep{navarro96} with a concentration index of $c=9.4$~\citep{lokas03}. The ICM density is computed using a $\beta$-profile with parameters derived from the Coma cluster~\citep{mohr99,fossati12}. This simple calculation suggests that, with the pericenter velocity of $4000\,{\rm km\,s^{-1}}$, D100 is likely to have been exposed to a ram pressure that is comparable or stronger than that in \rich\ in the last $\sim200\,$Myr.}.
However, note that the molecular hydrogen mass ($4.8\times10^8\,\msun$) and star formation rates ($2.3\,\msunyr$) of the D100 disk are larger by a factor 5--6 than those of the \rich\ galaxy~\citep{jachym17}. This indicates that D100 might have been even more gas-rich than the \rich\ galaxy before falling into the Coma cluster.

We also compare the properties of the RPS tails of D100 and the \rich\ galaxy. H$\alpha$ luminosity suggests that the total SFR in the tail of D100 is ${\rm d}M_*/{\rm d}t=3.9\times10^{-3}\,\msunyr$ \citep{jachym17}, which is  larger than the averaged SFR obtained in the simulated tail of the \rich\ galaxy (${\rm d}M_{\star}/{\rm d}t=1.2\times10^{-3}\,\msunyr$) at $t=135$--$366\,$Myr. 
The difference of a factor of three in the SFRs may be attributed to different gas masses in the tail. Although the observations of the D100 tail reveal a considerable amount of molecular hydrogen~\citep[$M_{\rm H_2}\sim10^9\,\msun$,][]{jachym17}, the observed tail is found to be HI-deficient~\citep{bravo-alfaro00,bravo-alfaro01}. In contrast, the tail of the \rich\ galaxy comprises a molecular hydrogen mass of $1.3\times10^8\,\msun$ with $M_{\rm H_2}/M_{\rm HI}\sim0.1$ at $t=366\,$Myr. Interestingly, $M_{\rm H_2}/M_{\rm HI}$ steadily increases over time in \rich, due to the rapid increase in H$_2$, compared to the increase of neutral hydrogen. Yet, the extreme H$_2$/HI ratio in the D100 tail is still difficult to explain based on the results of our simulations. \citet{jachym19} also showed that ESO137-001 in the Norma cluster has a H$_2$/HI ratio higher than unity, necessitating numerical studies of RPS galaxies with extremely high gas fractions.

\section{H$\alpha$ Emission from an RPS Galaxy}

A young stellar population emits the Lyman continuum (LyC) photons that ionize the surrounding gas. The ionized hydrogen subsequently recombines with electrons, producing H$\alpha$ photons at 6562.8\,${\rm \r A}$. Therefore, H$\alpha$ detection in RPS galaxies is often considered as an indication of star formation~\citep[e.g.][]{sheen17,yagi17,jachym17,jachym19}. However, H$\alpha$ photons can also be produced by collisional radiation, which does not require Lyman continuum radiation from young stars. Furthermore, heating due to processes other than star formation, such as shocks or mixing, can yield H$\alpha$. Thus, in this section, we investigate the origin of H$\alpha$ emission in RPS galaxies and discuss a possible way to determine star formation rates from H$\alpha$ in the tail.

Figure~\ref{fig:halpha} shows the dust-obscured H$\alpha$ SB maps of the gas-rich galaxy (\rich) at $t=185$ and 366\,Myr and the distribution of all disk stars (yellow contours), disk stars younger than 20\,Myr (blue contours), and tail stars older (open gray circles) and younger (open white stars) than $20\,$Myr. As strong ICM winds truncate the outskirts of the gaseous disk, the star-forming region notably shrinks in the disk between the two epochs. At $t=185\,$Myr, the stellar disk is bowed due to the gravitational interaction with the gaseous disk that is pushed by the strong ram pressure. However, the stellar disk recovers its shape once the gaseous disk is largely stripped, as seen at $t=366~\myr$. 

\begin{figure}
\centering 
\includegraphics[width=\linewidth]{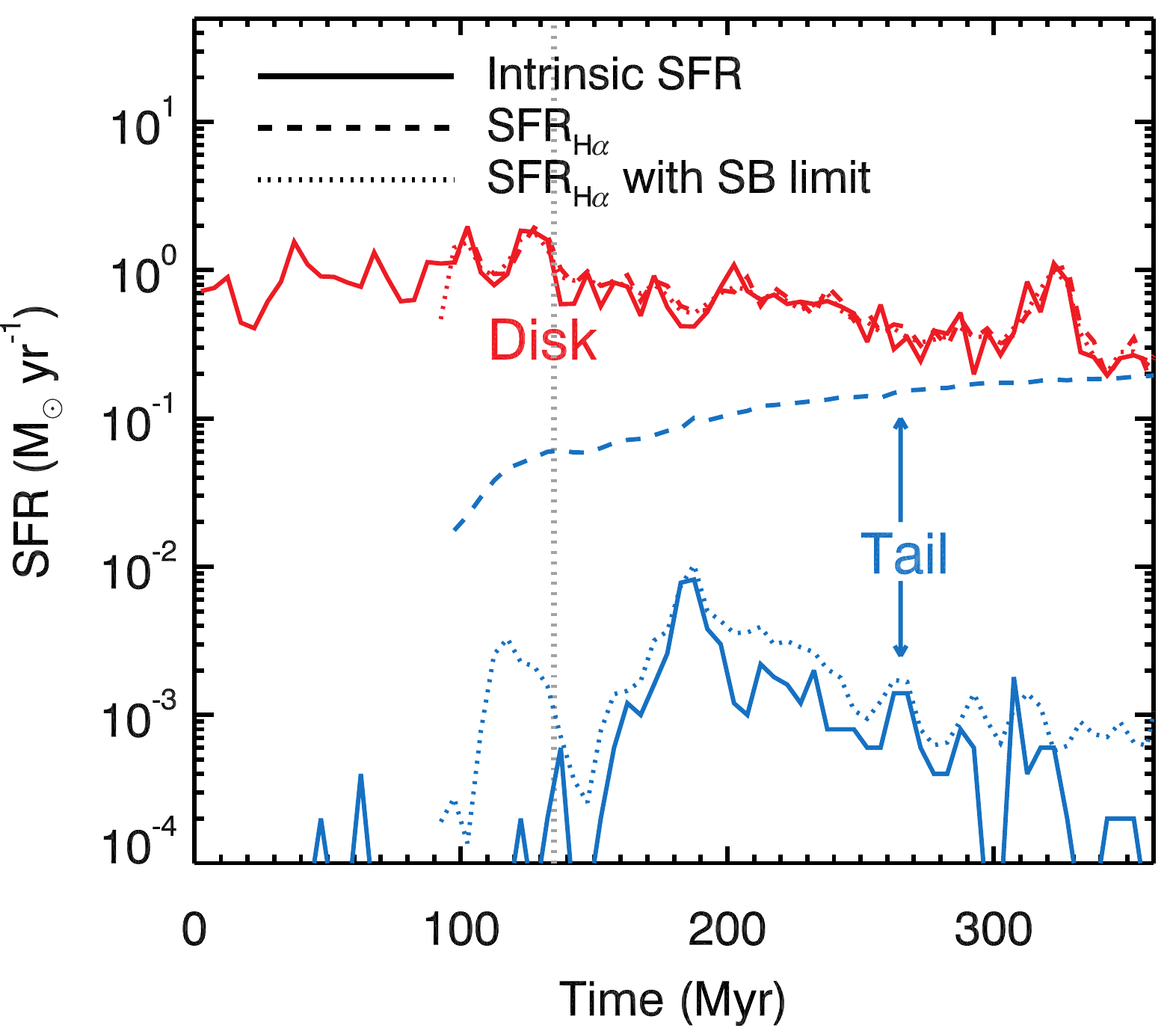}
\caption{Evolution of the intrinsic star formation rate (SFR, solid) and the SFR estimated from the total unobscured H$\alpha$ luminosity (SFR$_{\rm H\alpha}$, dashed and dotted). The red and blue colors indicate the SFRs in the disk and tail, respectively. We present SFR$_{\rm H\alpha}$ with (dotted) and without (dashed) a SB limit of $6\times10^{38}\,{\rm erg\,s^{-1}\,kpc^{-2}}$. We display the SFR$_{\rm H\alpha}$ lines only at $t>100\,$Myr. The vertical dotted line marks the epoch at which the wind front reaches the galaxy. SFR$_{\rm H\alpha}$ closely traces the intrinsic SFR in the disk, while the majority of the H$\alpha$ SB (dashed) in the tail originates from processes other than star formation. However, the intrinsic SFR in the tail is reasonably recovered by considering H$\alpha$ bright regions (dotted).}
\label{fig:sfr_halpha}
\end{figure}

Most young tail stars coincide with the local H$\alpha$ maxima. At $t=185\myr$, when the star formation rate in the tail peaks (Figure~\ref{fig:sfr}), a clump of stellar particles of $M_{\star}=7.2\times10^4\,\msun$ is born in the narrow region located at the interface between the tail and ICM wind, forming bright H$\alpha$ cores inside the white star symbols marked by \texttt{D} in Figure~\ref{fig:halpha}. Among them, the brightest $\rm H\alpha$ core has intrinsic $F_{\rm H\alpha}=2.41\times10^{41}\,{\rm erg\,s^{-1}\,kpc^{-2}}$ and its dust-obscured SB is $F_{\rm H\alpha}=2.11\times10^{38}\,{\rm erg\,s^{-1}\,kpc^{-2}}$. We confirm that the bright H$\alpha$ pixels are formed at the position of newly formed stellar particles. In the last stage of the simulation ($t=366\myr$), stellar particles younger than 20 Myr are observed across the entire tail with a total mass of $M_{\star}=1.8\times10^4\,\msun$. The brightest H$\alpha$ core in the tail has intrinsic H$\alpha$ SB of $6.15\times10^{40}\,{\rm erg\,s^{-1}\,kpc^{-2}}$ at $t=366\,$Myr, which is only a quarter of the brightest H$\alpha$ core in the tail at $t=185\myr$.

In Figure~\ref{fig:sfr_halpha}, we compare the intrinsic star formation rates averaged over 20 Myr and those estimated from the total H$\alpha$ luminosities in the disk and tail. To make the comparison from an observational perspective, we assume that the simulated galaxy lies at $z=0.0173$ and is observed by the Multi Unit Spectroscopic Explorer (MUSE) instrument on the Very Large Telescope, with a pixel scale of $0.2\,{\rm arc\,sec}$. Under these conditions, each pixel has a physical scale of $0.073\,{\rm kpc}$, assuming cosmological parameters $H_0=67.74\,{\rm km\, s^{-1}\,Mpc^{-1}}$ and $\Omega_0=0.3089$~\citep{planck15}. The H$\alpha$ luminosity of the disk is measured from the pixels covering the cylindrical volume of the disk with a radius of $r=12\,$kpc and a height of $h=\pm3\,$kpc from the galactic plane, as defined in \S 2.2. The H$\alpha$ luminosity of the tail is similarly obtained from the pixels covering the cylindrical volume extending from the upper surface of the disk to the boundary. Then, we infer the empirical star formation rate ${\rm SFR}_{\rm H\alpha}$ from the H$\alpha$ SB using a simple scaling relation between the star formation rate and an unobscured ${\rm H\alpha}$ luminosity: 
 \begin{equation}
 {\rm SFR}_{\rm H\alpha}\approx 4.0\, \msunyr \, \left( \frac{L_{\rm H\alpha}}{10^{42} \,{\rm erg\,s^{-1}}} \right),
 \label{eq:halpha}
 \end{equation}
 which is appropriate for a stellar population with a metallicity of $Z=0.01$ in \bpass\ \citep[v2.0,][]{eldridge08,stanway16}. 
 
 \begin{figure}
\centering 
\includegraphics[width=\linewidth]{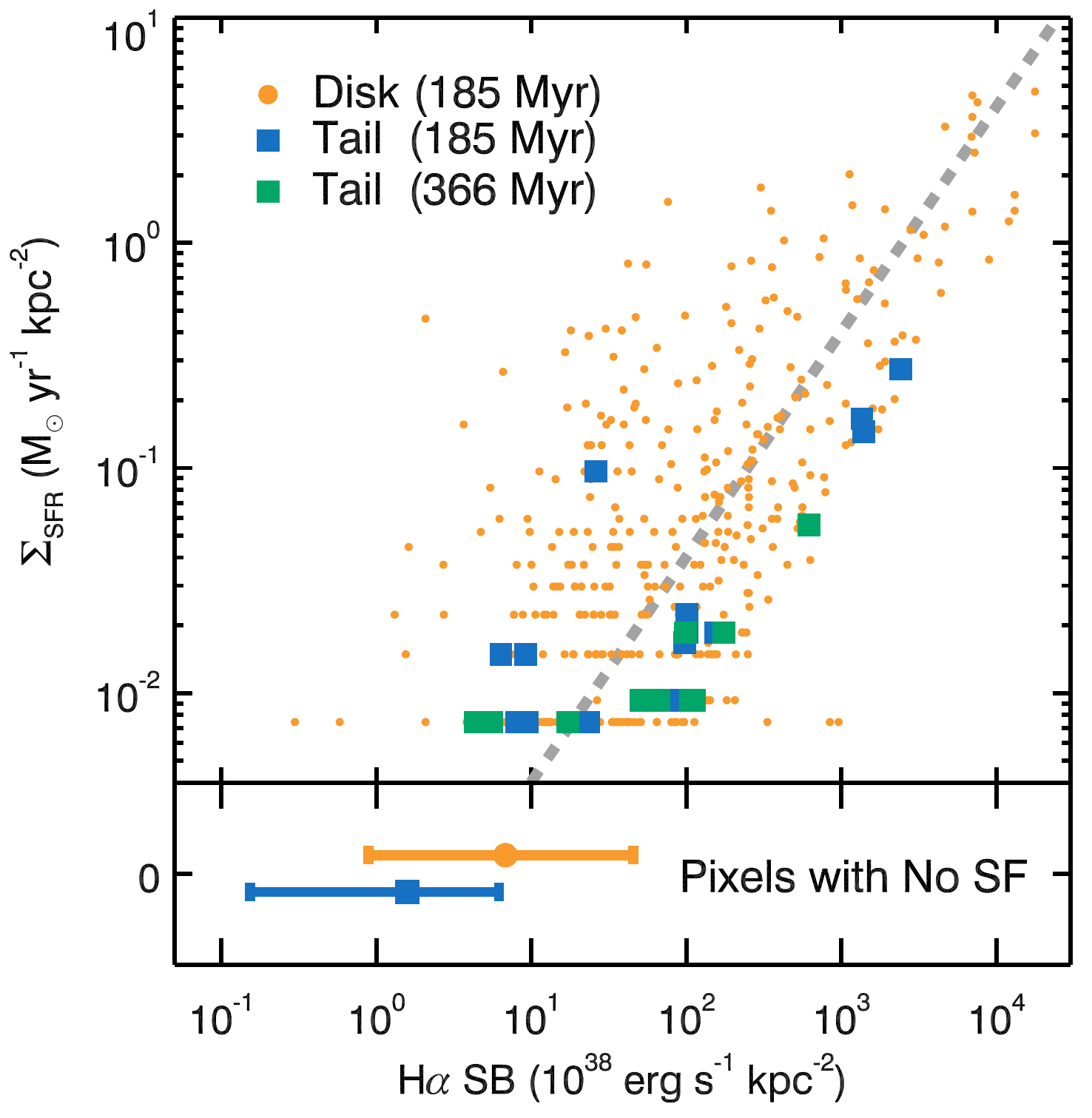}
\caption{The relation between the SFR surface density ($\Sigma_{\rm SFR}$) and H$\alpha$ SB  at $t=185\,$Myr when the SFR in the tail is maximal. The disk and tail regions are denoted by orange circles and blue squares, respectively. The green squares display the relation in the tail at the final snapshot ($t=366\,$Myr). We illustrate the SFR--H$\alpha$ relation expected from \bpass\ \citep{eldridge08,stanway16} with a gray dashed line. 
The colored bars in the bottom panel indicate the SB ranges of the pixels contributing to the $1^{\rm st}-99^{\rm th}$ percentile distribution of the total H$\alpha$ luminosities of the regions with no star formation in the tail (blue) and disk (orange). The orange circle and blue square in the colored bars mark the median values. Note that processes other than star formation lead to H$\alpha$ SB of up to $6\times 10^{38}\,{\rm erg\,s^{-1}\,kpc^{-2}}$ in the tail, indicating that the pixels with higher H$\alpha$ SB are likely to be observed as star-forming regions. }
\label{fig:halpha_sfr}
\end{figure}

  \begin{figure*}
\centering 
\includegraphics[width=1\textwidth]{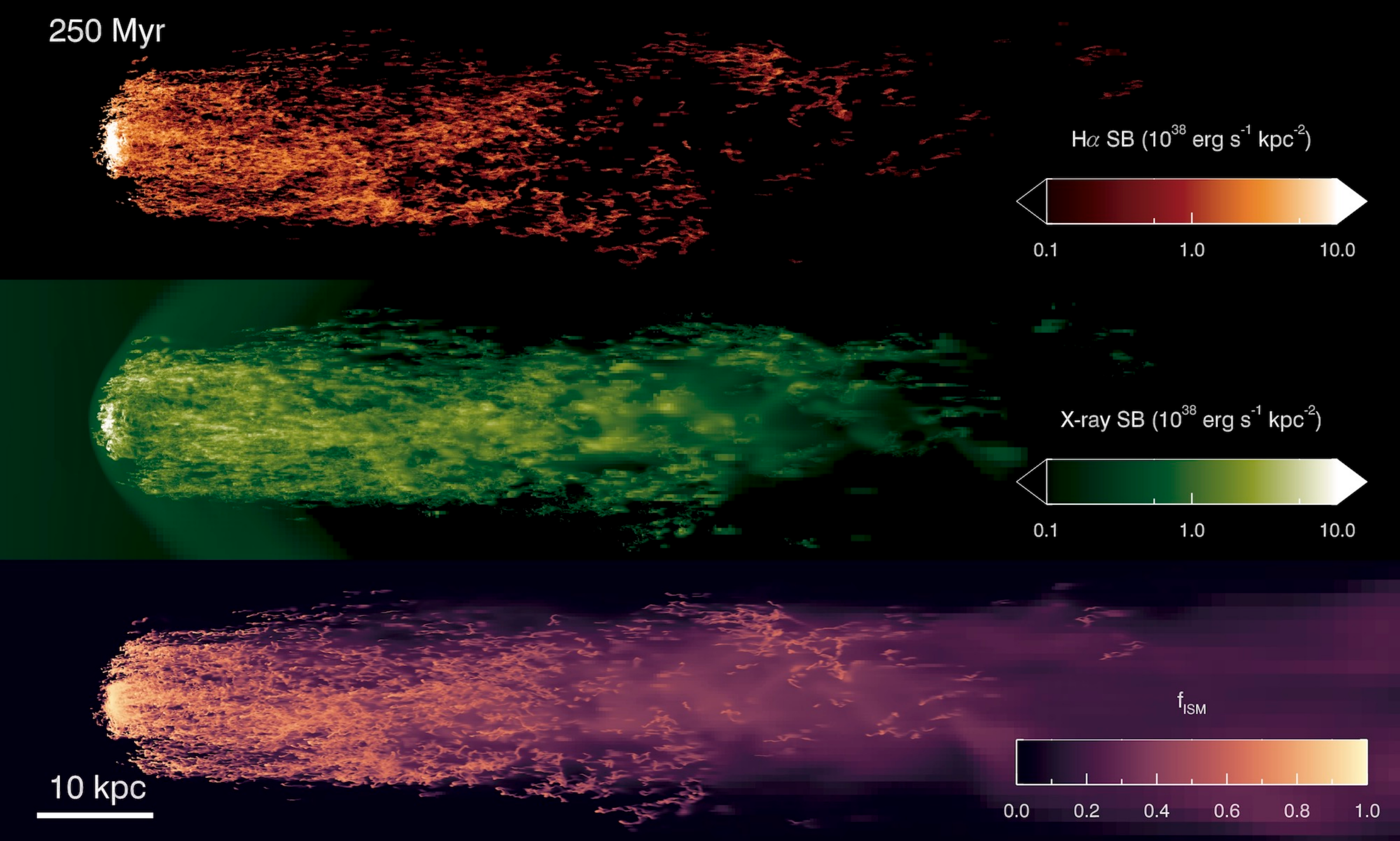}
\caption{Dust-obscured H$\alpha$ SB (top), X-ray SB in the energy band [0.4, 7.5] keV (i.e. before the bolometric correction, middle), and ISM fraction of the gas-rich galaxy at $t=250\,{\rm Myr}$ when the length of the tail is comparable to that of ESO137-001 and D100.}
\label{fig:img_xray}
\end{figure*}

Figure~\ref{fig:sfr_halpha} shows that, in the disk, SFR$_{\rm H\alpha}$ closely matches the intrinsic SFR within $\sim20\%$ error, regardless of the assumed SB limit. On the other hand, a marked difference is observed between the intrinsic SFR and ${\rm SFR}_{\rm H\alpha}$ in the tail. To determine the cause of the difference, we compute the relative contribution from collisional and recombinative transitions to the total number of H$\alpha$ photons. We find that $\approx 90\%$ of H$\alpha$ photons in the tail are emitted from the gas with $0.01<n_{\rm H}<1\,\cmq$, while  $\approx 10\%$ arises from the gas with $n_{\rm H}>1\,\cmq$ during the entire time period. At $n_{\rm H}>1\,\cmq$, more than 97\% of H$\alpha$ photons are emitted via recombinative transition, but in the diffuse gas, collisional radiation produces twice more H$\alpha$ photons than recombinative transition. Thus, we conclude that H$\alpha$ in the tail is mainly powered by a process unrelated to star formation, i.e. collisional radiation ($63\%$).  The remaining H$\alpha$ (37\%) is also unlikely to originate directly from star formation, given that the intrinsic SFR is several orders of magnitude smaller than ${\rm SFR}_{\rm H\alpha}$. Instead, we argue that the additional $\rm H\alpha$ photons in the tail arise from the interaction with a hot ICM which heats up the diffuse gas and stimulates the recombinative as well as collisional radiation.

In principle, Lyman continuum photons that manage to escape from the galactic disk can also contribute to $\rm H\alpha$ flux in the tail. However, we confirm that the LyC flux measured in the distant tail ($z>10\,$kpc) is insufficient  ($\lesssim15\%$) to explain the entire {\rm H$\alpha$} flux. We also estimate a possible contribution from the UV background radiation by measuring H$\alpha$ emission from low-density gas that is not self-shielded from the UV~\citep[$n_{\rm H}<0.01\,\cmq$,][]{rosdahl12}. We find that the H$\alpha$ photons emitted from the low density gas accounts for less than 0.5\% of the total $\rm H\alpha$ emission in the tail, and thus the UV background used in this study is also unlikely to power the ${\rm H\alpha}$ tails.

We further examine if bright H$\alpha$ blobs in the tail trace star forming regions by correlating an intrinsic star formation rate and unobscured H$\alpha$ SB in each pixel (which corresponds to 73 by 73 pc$^2$ in a physical scale) in Figure~\ref{fig:halpha_sfr}. The H$\alpha$ SB of each pixel from the disk (orange) and tail regions (blue) is projected at $t=185\myr$, when the star formation rate peaks in the tail. Although H$\alpha$ SB is notably scattered at a fixed $\Sigma_{\rm SFR}$, it follows reasonably well the predicted SFR-$\rm H\alpha$ relation (gray dashed line and Equation~\ref{eq:halpha}).

 For comparison, we present the SB range of the pixels contributing to the $1^{\rm st}-99^{\rm th}$ percentile distribution of the total H$\alpha$ luminosities of the regions with no stars younger than 20\,Myr, in the bottom panel.  Non-star forming regions in the disk can display pixels as bright as $10^{39}\,{\rm erg\,s^{-1}\,kpc^{-2}}$. However, the pixels in the tail with no star formation exhibit H$\alpha$ SB lower than $6\times10^{38}\,{\rm erg\,s^{-1}\,kpc^{-2}}$, which roughly corresponds to the typical H$\alpha$ SB at the minimum $\Sigma_{\rm SFR}$ of $\sim0.005\,\msun\,{\rm yr^{-1}} \,{\rm kpc^{-2}}$. Thus, we conclude that the bright ${\rm H\alpha}$ blobs in the tail are lit by nearby young stars. Additionally, our experiments suggest that the detection of the ${\rm H\alpha}$ emission brighter than $6\times10^{38}\,{\rm erg\,s^{-1}\,kpc^{-2}}$ can be considered as a sign of star formation in the RPS tail, while the well-developed tail structures with H$\alpha$ SB$< 10^{38}\,{\rm erg\,s^{-1}\,kpc^{-2}}$, observed in e.g., ESO 137-001 \citep{fumagalli14}, are likely to be induced by processes other than star formation. Thus, the discrepancy between the intrinsic SFR and ${\rm SFR}_{\rm H\alpha}$ in the tail is alleviated if a high SB limit is applied to include actual star-forming sites (e.g., $6\times10^{38}\,{\rm erg\,s^{-1}\,kpc^{-2}}$), as shown by the dotted line in  Figure~\ref{fig:sfr_halpha}.

\section{X-ray to H$\alpha$ flux ratio and mixing}

\begin{figure}
\centering 
\includegraphics[width=0.475\textwidth]{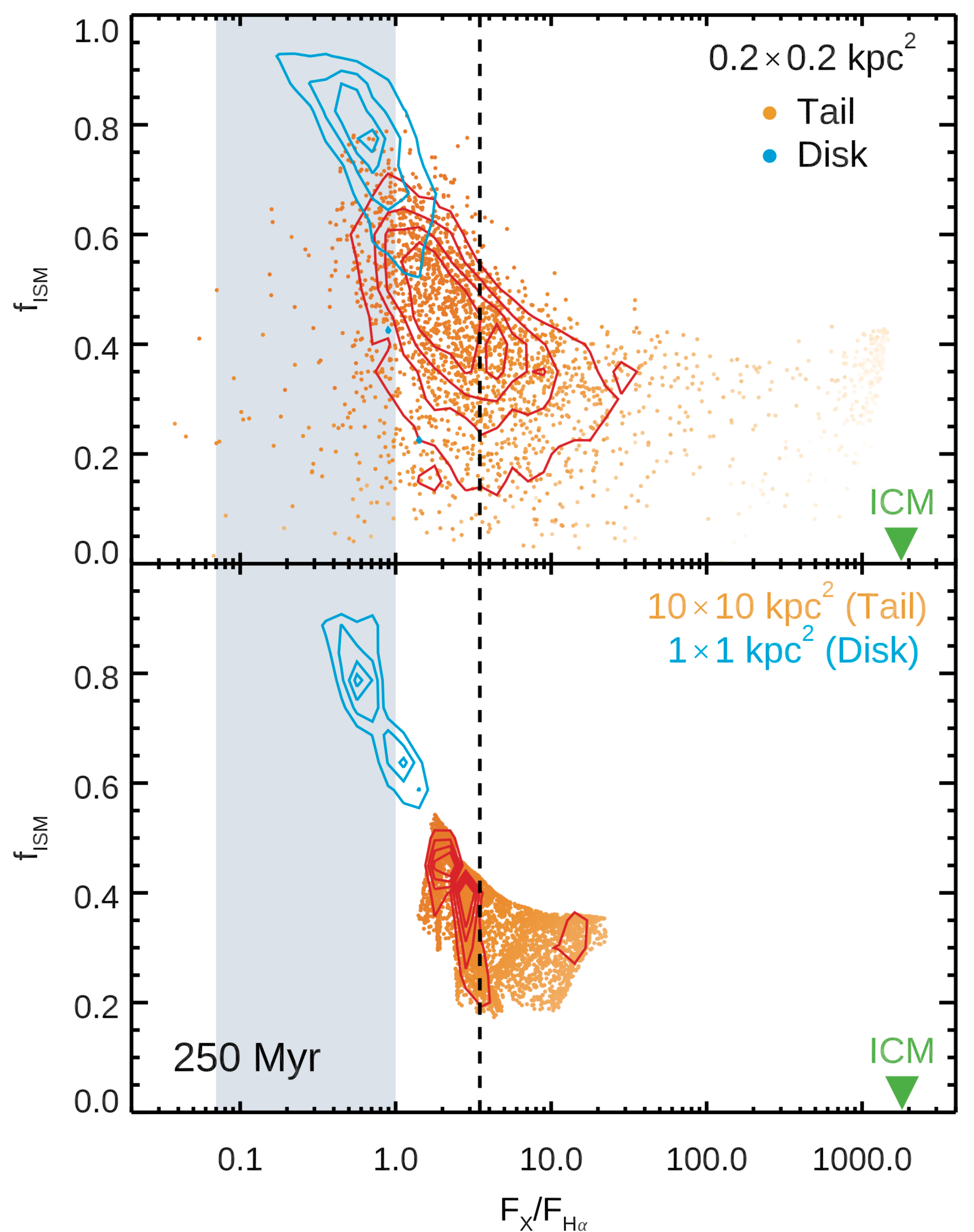}
\caption{ The correlation between $f_{\rm ISM}$ and the flux ratio of X-ray to dust-obscured H$\alpha$ in the disk (cyan contours) and RPS tail (orange dots and red contours) of the \rich\ galaxy at $t=250\myr$. The flux ratio of the imposed ICM wind is shown as green triangles. Darker orange dots correspond to the tail regions with brighter H$\alpha$. We compute the SB in pixels with area $0.2\times0.2\,{\rm kpc^2}$ in the upper panel, while larger pixels are used for the tail ($10\times10\,{\rm kpc^2}$) or for the disk region ($1\times1\,{\rm kpc^2}$) in the bottom panel. The light blue shaded regions and vertical dashed lines indicate the observed \SBR\ in the disk and RPS tail \citep{sun21}, respectively. 
}
\label{fig:fxh_fism}
\end{figure}

In addition to H$\alpha$ emission, RPS tails are often detected in X-rays~\citep{finoguenov04,wang04,machacek05,sun05,sun06,sun10,poggianti19}. Since X-ray photons are preferentially produced by hot plasma with $T\sim10^7\,$K, the co-existence of H$\alpha$ and X-ray suggests a complex thermal structure of the RPS tail. Furthermore, \citet{sun21} recently argued that the flux ratio of X-ray to H$\alpha$ (\SBR) may be used as a signature of the ICM--ISM interaction \citep[see also][]{poggianti19}. They showed that galactic disks typically have $\SBR\la 0.4$, but the RPS tails exhibit a high \SBR\ of 3--4, likely because of the mixing between the ICM and ISM~\citep[e.g.,][]{tonnesen11}.

To understand the mixing process and to gauge whether the H$\alpha$--X-ray flux ratio can be used as an indication of mixing, we generate mock X-ray SB maps of the RPS galaxy 
using the \yt\ package~\citep{turk11}. The X-ray emissivity is estimated as 
\begin{equation}
\epsilon_{\rm X}=n_{\rm HII}n_{\rm e} \, \left[\Lambda_{\rm X,p}(T)+\Lambda_{\rm X,m}(T,Z)\right],
\end{equation}
where $\Lambda_{\rm X,p}$ and $\Lambda_{\rm X,m}$ are the X-ray cooling functions for primordial gas and metals, respectively, taken from the Astrophysical Plasma Emission Code \citep[APEC,][]{smith01}. Following the bolometric correction procedure by \citet{sun21}, we first compute the X-ray emissivity in the $[0.4,\,7.5]\rm\,keV$ energy band and derive the bolometric flux using conversion factors measured from APEC as a function of temperature. The model bolometric correction factor at temperature $T$ is defined as $c_{\rm bol}(T)\equiv L_{\rm X,bol}(T)/L_{\rm X,[0.4,\,7.5]keV}(T)$, where $L_{\rm X,bol}(T)$ is the model bolometric luminosity and $L_{\rm X,[0.4,\,7.5] \,keV}(T)$ is the model luminosity at $[0.4,\,7.5]\,{\rm keV}$. X-ray emissivity-weighted temperature is used in the calculation of the correction factor.
Because we mimic the ICM wind by imposing a boundary condition of high temperature ($T=10^7\,$K), the ICM wind itself produces an X-ray background of $\sim5\times10^{37}\,{\rm erg\,s^{-1}\,kpc^{-2}}$. The background X-ray emission is removed by subtracting the X-ray SB profiles averaged at the boundary of the simulation box, similar to  observations. 

Figure~\ref{fig:img_xray} exhibits the maps of dust-obscured H$\alpha$ SB (top), X-ray SB (middle), and the mass fraction of the gas originating from the ISM ($f_{\rm ISM}$, bottom) in the gas-rich galaxy at $t=250\,$Myr. We select this snapshot because the simulated RPS tail extends out to $\sim 80\,{\rm kpc}$, which is comparable to that of ESO137-001 in the Norma cluster~\citep{sun21} and D100 in the Coma cluster~\citep{jachym17}. The X-ray SB and ISM fraction maps clearly indicate the presence of hot diffuse clouds mixed with the stripped ISM in the tail ($T\sim{\rm 10^6\, K}$ and $\nH\sim 0.01\,\cmq$). Such gas is bright in X-ray but is faint in H$\alpha$ because of the high temperatures (see e.g, the tail region at 30--40 kpc in Figure~\ref{fig:img_xray}). Conversely, H$\alpha$ bright clouds in the tails at $z>10\,{\rm kpc}$ are cooler ($T\sim 10^4\,{\rm K}$)  and denser ($n_{\rm H}\sim0.1\,\cmq$), and their contribution to the total X-ray luminosity is minor.

To understand the relationship between the flux ratio (\SBR) and the mixing, we sample the SB maps in the \rich\ run at $t=250\,{\rm Myr}$ with $0.2\times0.2\,{\rm kpc^2}$ pixels (upper panel) or $10\times10\,{\rm kpc^2}$ pixels (lower panel) in Figure~\ref{fig:fxh_fism}. The measurements on $1\times1\,{\rm kpc^2}$ pixels for the galactic disk at $|z|<3\,$kpc are also presented as blue contours in the lower panel. The black dashed line (\SBR=3.48) corresponds to the empirical fit to the 16 observed tails of the Virgo, Coma, A1367, and A3627 cluster galaxies~\citep[see][for further details]{sun21}. The light blue shade marks the approximate range of the empirical SB ratio of the galactic disk also taken from \citet{sun21}. Since no extinction corrections are applied to the H$\alpha$ SB of the diffuse tails in \citet{sun21}, we also use the dust-obscured H$\alpha$ SB maps\footnote{However, we note that attenuation due to dust in our simulated tail is negligible, because the tail gas is highly ionized and a large amount of dust is destroyed (see Eq.~\ref{eq:dust}).}.

We find that the galactic disks form a distinctive sequence in Figure~\ref{fig:fxh_fism}, with the flux ratio $\SBR\la$ 1.5, consistent with the observations~\citep{sun21}. The mean flux ratio of the simulated disk tends to be slightly larger than the average observed ratio, which is likely due to the boosted SN feedback employed in this work.  In contrast, the pure ICM gas shows a higher \SBR\ of $\sim$ 1800  (the green triangles in Figure~\ref{fig:fxh_fism}). The mixing of the ISM into the ICM leads to the decrease of the hot gas temperature, and thus X-ray emissivity is reduced while ${\rm H\alpha}$ becomes brighter with increasing $f_{\rm ISM}$. Consequently, the ICM-dominant tail gas shows $\SBR\sim 1$--$100$, while the gas in the disk vicinity ($3<z<10\,{\rm kpc}$) exhibits a lower $\SBR$ of $\sim$1. This is again compatible with the observed trend that \SBR\ tends to increase for distant tails \citep{sun21}. Therefore, we argue that the intermediate SB ratio observed in the RPS tail can be seen as a sign of ICM--ISM mixing \citep{tonnesen11}.

We also remark that the predicted flux ratios in the tail ($\SBR\sim$ 1.5--3.5) are slightly smaller than those observed~\citep[$\SBR=3.48$,][]{sun21}. Given that a large amount of dust in the relatively hot tail gas is unlikely to survive and reduce $F_{\rm H\alpha}$, it is more probable that the X-ray fluxes in the tail are under-estimated. The correlation between $f_{\rm ISM}$ and \SBR\ in Figure~\ref{fig:fxh_fism} then suggests that the ICM should be mixed with the ISM more efficiently in the tail. This may be achieved by including thermal conduction \citep[e.g.,][]{armillotta17,li20} and/or by resolving hydrodynamic instabilities with higher resolution.  

Finally, it is also worth pointing out that the flux ratios in the simulated tail are widely distributed on small scales (0.1 kpc), while they tend to converge to $\sim1$ when sampled on kpc scales. This suggests that the difference in the flux ratio between the disk and RPS tails found by \citet{sun21} may be less dramatic if high-resolution observational data are obtained.

\section{Conclusions}

We investigated the formation of jellyfish galaxies using a set of idealized simulations for a dwarf-sized galaxy with a multi-phase ISM in environments with and without strong ICM winds devised to mimic the ram pressure at a cluster center. As a follow-up study of \citetalias{lee20}, we adopted the same code, physics, simulation setup, and initial condition to those of \citetalias{lee20}, but with a raised initial gas fraction. We primarily focused on the formation process of multi-phase clouds and stars in RPS tails. We found that the mixing of a stripped ISM with the ICM is a key process determining the characteristics of a jellyfish galaxy. Our results are summarized as follows.

\begin {enumerate}

\item 
Strong ram pressure efficiently suppresses star formation in the disk of a gas-rich galaxy by truncating the outskirts of the gaseous disk, which is consistent with the prediction of \citet{gunn72}. The decaying trends in star formation rates in the galactic disk are similar for the \rich\ and \normal\ galaxies, in spite of their different initial gas fractions ($M_{\rm HI}/M_\star= 4.1$ vs. $0.8$). 

\item
Molecular gas can form {\it in-situ} in the distant RPS tail of the \rich\ galaxy. The stripped ISM is mixed with the ICM, enhancing the formation of warm ionized gas in the RPS tail. Half of the HII clouds with $f_{\rm ICM}>0.5$ have cooling timescales shorter than a few Myr in \rich, which is in contrast to the results of the \normal\ galaxy, because of the lack of stripped ISM. This indicates that the stripping of a large amount of ISM plays a critical role in the formation of molecular clouds in RPS tails.

\item
 The RPS tails in the \rich\ galaxy form stars at a rate of $1.2\times10^{-3}\,\msunyr$ on average after the galaxy encounters the ICM wind.  The majority of the tail stars are initially formed in the stripped wake within $10\,$kpc from the galactic plane, but a small amount of stars ($8\times10^3\,\msun$) also forms in the distant tail ($z>60$ kpc) $\sim$200\,Myr after the galaxy starts to interact with the ICM wind. Stars in the distant tail form out of molecular clumps that are comprised of gas which is a mixture of stripped ISM and the ambient ICM ($f_{\rm ISM}\sim0.5)$.

\item 
The intrinsic star formation rate in the disk is reasonably recovered from the H$\alpha$ emission. In the RPS tail, only bright H$\alpha$ cores trace actual star-forming regions. H$\alpha$ emission below $6\times10^{38}\,{\rm erg\,s^{-1}\,kpc^{-2}}$ can originate from processes other than star formation, and the high SB limit needs to be imposed to recover the intrinsic SFR from $\rm H\alpha$ in the simulated RPS tail.

\item 
 A strong correlation is present between the ISM fraction and the flux ratio of X-ray to H$\alpha$ (\SBR) in \rich. The typical flux ratio in the RPS tail ($1.5\la \SBR \la 20$) is higher than that of the galactic disk ($\SBR\la$1.5) and lower than that of the ICM gas ($\SBR\sim$1800) when measured on $10\times10\,{\rm kpc^2}$ scales. 
 Although a factor of two difference is seen between the tail \SBR\ of our model and the empirical fit of \citet{sun21}, the trend still clearly supports the interpretation that the observed intermediate flux ratio indicates the mixing between the ISM and ICM.

\end {enumerate}

We have shown that several RPS features can be reproduced when a large amount of ISM material is stripped from a gas-rich galaxy via strong ram pressure, forming prominent multi-phase tails due to mixing with a hot ambient medium. However, several issues still need to be addressed in future studies. First, thermal conduction is not included in this work. \citet{li20} examine the effects of radiative cooling, self-shielding, self-gravity, magnetic field, and Braginskii conduction and viscosity, and show that cooling and conduction are the physical processes that govern the lifetime of cool clouds in a circumgalactic medium (CGM). In their study, conduction efficiently evaporates small clouds that are weakly or not at all self-shielded while it hardly affects cool and dense clumps. This suggests that conduction could suppress the growth of cold clumps in the RPS tail. Second, our simulations do not include explicit viscosity. \citet{roediger15b} demonstrate that the mixing becomes less efficient with increasing viscosity in the RPS tails. \citet{li20} also show that a boundary layer formed by viscosity can insulate the stripped ISM from the ICM, increasing the lifetime of the cool clouds. Although it is clear that viscosity plays a role in the mixing process, we note that strong viscosity effects would suppress X-ray emissivity in the tail, potentially aggravating the agreement seen in $\SBR$ (Figure~\ref{fig:fxh_fism}). If this is really the case, the viscosity effects should be offset by other processes, such as conduction~\citep{li20}. Third, we assume a constant ICM wind in this work, but galaxies orbiting around the real cluster would undergo ram pressure that changes over time \citep[e.g.,][]{roediger07,roediger15a,jung18,yun19}. For example, \citet{tonnesen19} shows that the gas stripping rates and tail sizes become smaller if the wind strengths are gradually increased, compared to the run with constant winds. In this regard, the impact of the ram pressure in our simulations may be overestimated, perhaps enhancing the star formation in the RPS tail. Fourth, we neglect UV radiation from AGNs or star-forming BCGs~\citep{hicks10,klesman12,fogarty15,poggianti17} in a cluster environment. These sources can provide extra heating, potentially limiting star formation in the RPS tail. Unfortunately, the effect of the stronger UV radiation depends on (variable) AGN or star formation activity, which is difficult to estimate without performing realistic simulations. Last but not the least, our simulations still do not explain some key properties of observed RPS tails. For example, the D100 tail in the Coma cluster is observed to have star formation rates comparable with or only a factor of a few higher than that of the \rich\ galaxy, but the amount of molecular clouds is ten times larger than that in \rich. Even more intriguing is the HI deficiency in the D100 and ESO137-001 tails \citep{jachym14,jachym17,jachym19}, indicating that RPS gas may turn into molecular clouds very efficiently in certain conditions. 
These issues necessitate future studies probing a larger parameter space and physical ingredients in realistic cluster environments.

\section*{acknowledgments}
The authors would like to thank the anonymous referee for their constructive review of this manuscript. The authors also thank Ming Sun for helpful comments on the comparison of X-ray luminosities.
JL is supported by the National Research Foundation of Korea (NRF-2021R1C1C2011626). TK is supported by the National Research Foundation of Korea (NRF-2020R1C1C1007079), and acted as the corresponding author. YKS acknowledges support from the National Research Foundation of Korea (NRF) grant funded by the Ministry of Science and ICT (NRF-2019R1C1C1010279). 
The supercomputing time for numerical simulations was kindly provided by KISTI (KSC-2020-CRE-0278), and large data transfer was supported by KREONET, which is managed and operated by KISTI. This work was also performed using the DiRAC Data Intensive service at Leicester, operated by the University of Leicester IT Services, which forms part of the STFC DiRAC HPC Facility (www.dirac.ac.uk). The equipment was funded by BEIS capital funding via STFC capital grants ST/K000373/1 and ST/R002363/1 and STFC DiRAC Operations grant ST/R001014/1. DiRAC is part of the National e-Infrastructure.

\end{document}